\documentclass[letterpaper, 10 pt, conference, article]{ieeeconf} 
\IEEEoverridecommandlockouts                              
\usepackage[a4paper,margin=1in,]{geometry}
\usepackage{tabularx}
\usepackage{ragged2e}
\usepackage{multirow}
\usepackage{colortbl}
\usepackage[table,xcdraw]{xcolor}
\usepackage{arydshln}
\usepackage[a4paper,margin=1in]{geometry}
\usepackage{graphicx}
\usepackage{booktabs}
\usepackage{tikz}
\usepackage[font=normal,labelfont=bf]{caption}
\usepackage{url}

\makeatletter
\let\NAT@parse\undefined
\makeatother

\definecolor{lightgray}{gray}{0.95}
\definecolor{lightblue}{HTML}{E6F0FA}

\definecolor{linkblue}{HTML}{1A4D8F}
\definecolor{iacrPink}{HTML}{FF0099}  

\definecolor{iacrPink}{HTML}{FF0099}
\usepackage[colorlinks=true,
            linkcolor=iacrPink,
            urlcolor=iacrPink,
            citecolor=iacrPink]{hyperref}

\usepackage[T1]{fontenc}
\usepackage{tgpagella}
\usepackage{authblk}

\usetikzlibrary{positioning,shapes,arrows.meta, fit, calc}

\title{\LARGE \bf 
\textbf{QORE : Quantum Secure 5G/B5G Core}
}

\author[1]{Vipin Rathi}
\author[2]{Lakshya Chopra}
\author[2]{Rudraksh Rawal}
\author[2]{Nitin Rajput}
\author[2]{Shiva Valia\\}
\author[2]{Madhav Aggarwal}
\author[2]{Aditya Gairola}

\affil[1]{Ramanujan College, University of Delhi, New Delhi India}
\affil[2]{coRAN Labs Private Limited, New Delhi, India}

\begin{document}
\maketitle
\thispagestyle{empty}
\pagestyle{empty}

\begin{abstract}
Quantum computing is rapidly reshaping the security landscape of modern telecommunications. The cryptographic foundations that secure today’s 5G systems—RSA, Elliptic Curve Cryptography (ECC), and Diffie–Hellman (DH)—are all susceptible to attacks made feasible by Shor’s algorithm. As a result, protecting 5G networks against future quantum adversaries has become an urgent research and engineering priority.

In this paper, we introduce QORE, a quantum-secure 5G/B5G Core framework that provides a clear pathway for transitioning both the 5G Core Network Functions and User Equipment (UE) to Post-Quantum Cryptography (PQC). The framework relies on the NIST-standardized lattice-based algorithms—Module-Lattice Key Encapsulation Mechanism (ML-KEM) and Module-Lattice Digital Signature Algorithm (ML-DSA)—and applies them across the 5G Service-Based Architecture (SBA). A Hybrid PQC (HPQC) configuration is also proposed, combining classical and quantum-safe primitives to maintain interoperability during migration.

Experimental validation indicates that ML-KEM delivers quantum security with only minor performance overhead, satisfying the stringent low-latency and high-throughput requirements of carrier-grade 5G systems. The proposed roadmap aligns with 3GPP SA3 and SA5 study activities on security and management of post-quantum networks, as well as NIST PQC ongoing standardization efforts and offers practical guidance for mitigating quantum-era risks while safeguarding long-term confidentiality and integrity of network data.

\end{abstract}

\section{Introduction}

For a long time, the telecommunication sector has evolved over time through innovation and introduced new possibilities. In the current times, the pace of evolution has increased at a high rate. But as time changes, there is a need for a new, better, and immediate security paradigm in 5G. There has been a dependence on classical Public Key Cryptography methods such as RSA, Diffie–Hellman (DH), and Elliptic Curve Cryptography (ECC) in 5G security systems, and even now, it still anchors them. However, advancements in quantum computing also create challenges for security systems due to the capability of quantum computers to execute Shor’s algorithm~\cite{nist-pqc-report}, which can break classical PKC. Every protocol in 3GPP relying on PKC for authentication or key exchange be it TLS, DTLS, ECIES, or IKEv2 faces the same challenge and makes 5G quantum vulnerable.

Authentication and Key Agreement in 5G use symmetric encryption, which, for the foreseeable future, remains robust against the advent of quantum computing. However, a single cryptographic breakthrough could expose a vast amount of data, breaking the entire system. The \textit{Harvest Now, Decrypt Later} \cite{nist_pqc_2024} strategy highlights this danger, where an adversary can collect encrypted data today and decrypt it later when the capabilities of quantum computing mature. Keeping this context in mind, a delay in migration to quantum-proof security is no less than a strategic failure.

Our contribution in this paper presents a migration framework for transitioning the 5G Core (5GC) security to quantum-resilient security. Our focus in this paper outlines securing inter-functional channels such as IPsec tunnels over the N2 and N3 interfaces~\cite{3gpp_33501} and TLS 1.3 protected Service Based Interfaces (SBI)~\cite{3gpp_23501}. The architecture incorporates National Institute of Standards and Technology (NIST) standardized lattice-based Post-Quantum Cryptography primitives namely, ML-KEM for key encapsulation and ML-DSA for digital signatures~\cite{fips203, fips204}. In the 5G system, efficiency plays a vital role, an important constraint in the architecture for real-time 5GC performance, and cannot be treated as an afterthought. For a smooth transition, we employ a hybrid model that combines traditional ECC with PQC elements such as ML-KEM, ML-DSA, and PQ authentication methods. Guided by RFC 8784 \cite{rfc8784} and several Internet drafts that are in progress, this design provides operational continuity, aligns with 3GPP SA3’s security roadmap, and offers a practical path to diminish quantum-era threats.

\section{Background}

With the introduction of technologies such as mmWave, enhanced Mobile Broadband (eMBB), ultra-Reliable Low-Latency Communication (uRLLC), and a microservice-based 5G Core architecture with cloud-native and edge deployments, 5G enables a wide range of new use cases. Naturally, this expansion increases the attack surface, requiring robust security mechanisms. 5G addresses these challenges through comprehensive frameworks for mutual authentication, signaling integrity, and the confidentiality of both user and control planes~\cite{3gpp_33501}.

These security enhancements diminish numerous current vulnerabilities, but they also bring along extra operational and architectural complexity. This leads to continuous observation to protect sensitive data and also preserve user anonymity.

Assuring 5G deployment security remains paramount not just for the resilience of network functionality but also for reliably protecting user data as it traverses the networks. In spite of these improvements, these long-standing cryptographic approaches utilized under 5G are nevertheless susceptible to those risks posed by the continued growth of quantum computing. Specifically, quantum algorithms like Shor's are capable of rapidly solving those complex mathematical problems upon which current systems like RSA and ECC are founded, and thereby render current encryption mechanisms obsolete. For this reason, 5G systems should adopt quantum-resistant techniques to ensure lasting data security. 

PQC provides cryptographic primitives of constructions (encryption and signature schemes) specifically designed to counter these risks, and consequently lays an essential foundation for enduring network integrity and confidentiality. Strategic integration of the latter, therefore, remains necessary for the sustainable and secure evolution of 5G infrastructure.

\subsection{Security Protocols}

\subsubsection{Transport Layer Security (TLS)}

Within the 5GC Service Based Architecture(SBA) communication, TLS provides integrity, privacy and authentication for a secure data transfer. TLS 1.3 employs both asymmetric cryptography including RSA or ECDSA and modern elliptic curve schemes such as X25519 for authentication and key establishment and symmetric encryption namely AES-GCM or ChaCha20-Poly1305 for the transfer of data. 

Mutual authentication is achieved in TLS through X.509 certificates, with Elliptic Curve Diffie-Hellman Ephemeral(ECDHE) ensuring Perfect Forward Secrecy (PFS). However, both ECDHE and classical signatures schemes are susceptible to quantum attacks, advising the  migration to secure, dependable and efficient PQ algorithms.


\subsubsection{Datagram Transport Layer Security (DTLS)}
Datagram TLS (DTLS) adapts TLS for UDP-based communication, securing latency-sensitive 5G control-plane exchanges such as N2 signaling between gNB and AMF. It retains TLS security features while handling packet loss and reordering via sequence numbers and retransmissions. DTLS inherits TLS’s quantum vulnerabilities and faces added challenges when integrating larger post-quantum keys due to UDP fragmentation. In 5G networks, DTLS is employed on inter-site and RAN interfaces \cite{oran-sec} (e.g., NGAP, F1AP) to securely encapsulate SCTP traffic \cite{rfc8261}, provide mutual authentication between O-CU, DU, and the core network, and protect application-layer messages.

\subsubsection{Mutual Transport Layer Security (mTLS)}
Mutual TLS (mTLS) enforces two-way authentication, where both client and server present X.509 certificates. Used extensively across 5G SBI interfaces, mTLS ensures NF-to-NF trust under a zero-trust model \cite{3gpp_33501,nist_sp800_207}. While resilient against impersonation and MITM attacks, its ECDSA/RSA and ECDHE primitives are quantum-vulnerable, motivating post-quantum upgrades such as ML-DSA and ML-KEM.

\subsubsection{OAuth 2.0}
OAuth 2.0 is a widely adopted authorization framework that enables third-party applications to securely access HTTP resources with limited permissions, either on behalf of the resource owner or for their own account \cite{rfc6749}.
In the 5G Core network. In the context of 5G Service-Based Architecture, OAuth 2.0 provides token-based authentication and authorization mechanisms for inter-NF communication. The framework defines four roles: Resource Owner (the NF owning the resource), Resource Server (the NF hosting protected resources), Client (the NF requesting access), and Authorization Server (the NF issuing access tokens, typically the Network Repository Function - NRF). The OAuth 2.0 flow involves the client NF requesting an access token from the authorization server (AS), which validates the request and issues a digitally signed JSON Web Token (JWT). After receiving the JWT from the AS, the client sends an access request (via HTTPS) to the resource server, including the access token in its payload, which the server then verifies. Access to the resource is granted on successful verification. The security of OAuth 2.0 in 5GC is critical and depends on classical cryptographic signatures, such as RSA and ECDSA, for protection against token tampering, and forgery.  The authorization server signs tokens using its private key, and resource servers verify these signatures using the corresponding public key. However, these signature schemes are vulnerable to quantum attacks, potentially allowing adversaries with quantum computers to forge authorization tokens and gain unauthorized access to network services. Therefore, migrating OAuth 2.0 to use post-quantum signature algorithms like ML-DSA is essential for maintaining authorization security in quantum-threatened environments.

\subsubsection{Internet Protocol Security (IPsec)}
IPsec is a set of open-standards for securing the network layer over public networks. IPsec provides high-quality cryptographic security for both IPv4 and IPv6. It offers data confidentiality, access control, integrity, authentication, traffic flow confidentiality, etc. In 5G Core networks, particularly for inter-site and user-plane traffic, it supports both Transport and Tunnel modes, enabling secure NF-to-NF or UPF-to-UPF data exchange over potentially untrusted IP networks. \cite{nist_sp800_77r1} provides comprehensive guidance for deploying highly secure and high-performance IPsec systems, emphasizing cryptographic configurations, throughput optimization, and interoperability testing. IPsec relies on core sub-protocols such as Encapsulating Security Payload (ESP) for confidentiality and Authentication Header (AH) for data integrity and authentication. Key management in IPsec is typically achieved through Internet Key Exchange version 2 (IKEv2), which uses Diffie–Hellman key exchange and digital certificates for mutual authentication \cite{rfc6071} \cite{rfc7296}. Similar to TLS, classical IKEv2 key exchange and signature algorithms (e.g., ECDH, RSA, ECDSA)\cite{nist-pqc-report} are vulnerable to quantum attacks, motivating the adoption of post-quantum key exchange and signature schemes to ensure long-term confidentiality and integrity of 5G communications.

\subsection{Preliminaries}

\subsubsection{Service-Based Architecture (SBI) and Core Network Functions}
The 5G Core (5GC) is fundamentally based on the Service-Based Architecture (SBI), utilizing virtualization for inter-functional communication between Network Functions (NFs) such as the Access and Mobility Management Function (AMF), Session Management Function (SMF), and Network Repository Function (NRF). This architecture usually relies on interfaces which are secured for session control, key management and mobility management. Relying heavily on certificates for authentication and establishing trust, typically secured via mTLS, the SBI facilitates communication between the NFs, 

\subsubsection{NF Communication and Cryptography}
Inter-NF communication over the SBI is secured using TLS, with digital signatures used in certificates and key exchange mechanisms (such as Elliptic Curve Diffie-Hellman, ECDH) forming the basis of the security perimeter. These digital signatures and key exchange protocols rely on classical asymmetric cryptography. The integrity of communication links between the 5GC and the Radio Access Network (RAN), specifically the N2 and N3 interfaces, are secured using IPsec/IKEv2, which similarly employs Diffie-Hellman or ECDH for key establishment.

\subsubsection{Authentication Mechanism (AKA)}
The Authentication and Key Agreement (AKA) Procedure is the standard process used by the 5GC to verify user identities and establish a secure session key. While the AKA procedure involves key generation, the core symmetric key protocols used for radio encryption and integrity—specifically the 128-bit symmetric algorithms—are generally considered secure for the foreseeable lifetime of 4G and 5G legacy systems, even with the arrival of quantum computers.

\subsubsection{Subscriber Identity Protection (SUPI to SUCI Conversion)}
Protecting subscriber identity is a high-priority security feature in 5G, achieved partly through the Subscription Concealed Identifier (SUCI). This mechanism replaces the vulnerable International Mobile Subscriber Identity (IMSI) by encrypting the Subscription Permanent Identifier (SUPI) into the SUCI before transmission. The Elliptic Curve Integrated Encryption Scheme (ECIES) is the current encryption method for this crucial privacy step. However, because ECIES is an Elliptic Curve-based public-key scheme, it is fundamentally vulnerable to quantum attacks, which immediately places long-lived subscriber identities at risk.

\section{Threat Analysis}

\begin{figure*}
    \centering
    \includegraphics[width=1.1\textwidth]{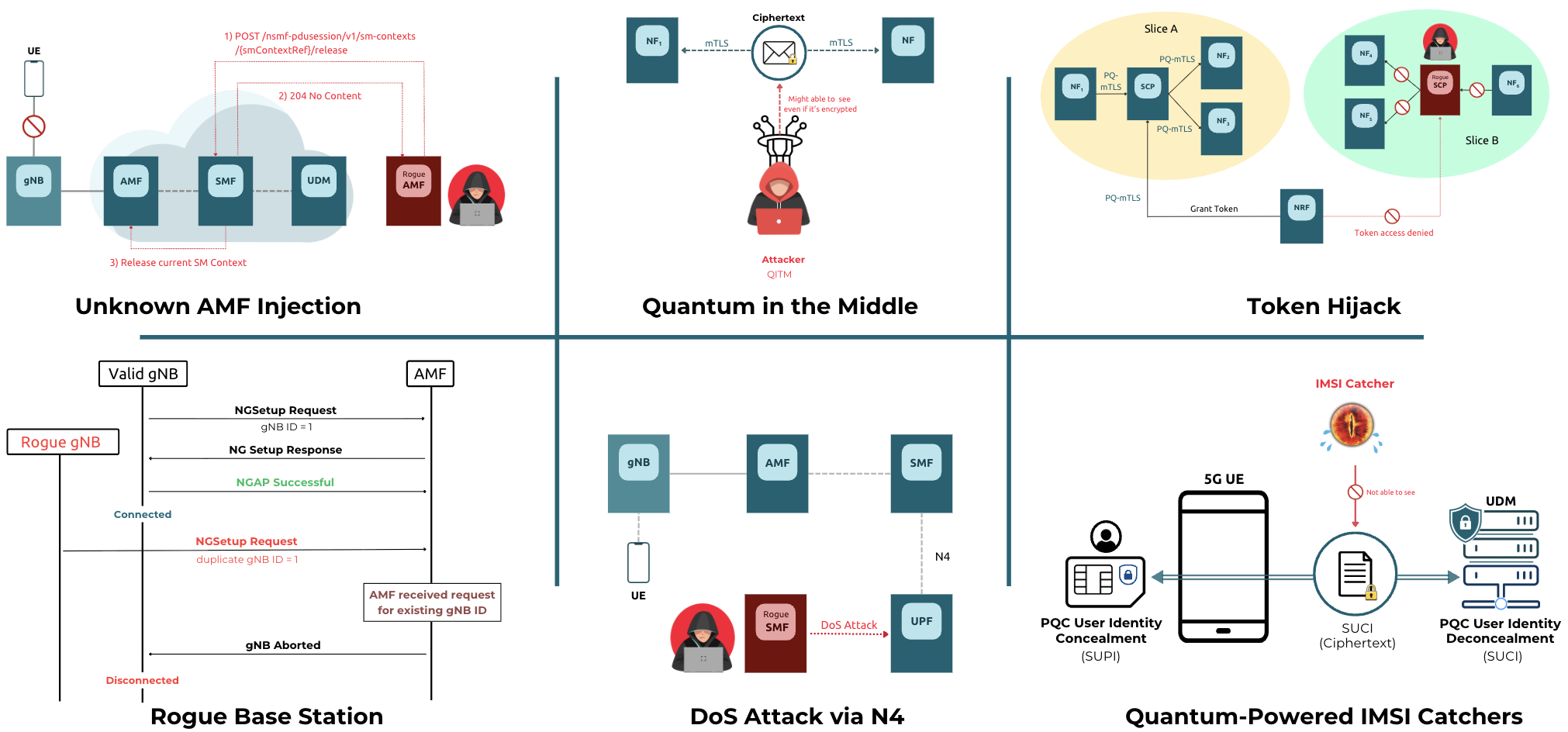}
    \caption{Security Threats to 5G Infrastructure from Quantum Computers. This diagram illustrates multiple quantum-enabled attack vectors affecting different components of the 5G architecture, including threats to UE authentication, token-based authorization, subscriber identity protection, and inter-NF secure communications.}
    \label{fig:threats}
\end{figure*}

The quantum threat to 5G infrastructure materializes through various attack vectors that target the cryptographic foundations of network security, with the comprehensive landscape illustrated in Figure~\ref{fig:threats} for a clearer prioritization of quantum-resistant security measures.

\subsection{Attack Taxonomy}

\subsubsection{Passive Attacks}

\textbf{Harvest Now, Decrypt Later (HNDL):} This represents the most insidious quantum threat to 5G networks. Adversaries intercept and store encrypted traffic today with the intention of decrypting it once sufficiently powerful quantum computers become available. This attack is particularly dangerous for long-lived sensitive data such as subscriber identities (SUPI), authentication credentials, and confidential communications that may retain value for years or decades. The ability of quantum-powered adversaries to retrospectively decrypt nearly all modern communications serves as a strong reminder to maintain crypto-agility in critical networks, including 5G. The HNDL threat affects all communications secured with classical PKC, including IPsec tunnels (N2/N3 interfaces), TLS-protected SBI communications, and ECIES-encrypted SUCI.

\textbf{Breach of confidentiality:} CRQCs running Shor's algorithm are the biggest threat to privacy because they can break RSA and ECC encryption directly, which means that anyone who intercepts communications can see the plaintext content. This flaw has a huge impact on session keys that are set up using classical Diffie-Hellman or ECDH key exchange in key protocols like TLS and IKEv2. It makes all data sent over those sessions completely compromised. Additionally, the breach affects other important parts of the network, such as OAuth tokens, certificate private keys, and pre-shared keys. This could cause the whole security infrastructure to fail at once.

\subsubsection{Active Attacks} 
Active attacks are those where adversaries actively attempt to compromise digital communications by having a sufficiently large quantum computer (CRQC) at their disposal. These attacks pose risks such as impersonation, man-in-the-middle, data compromise, and eavesdropping. Below, we describe how these attacks could impact 5G Core networks.

\noindent\paragraph*{Impersonation/Spoofing Attacks}

\textbf{\\Network-Level Spoofing:} An active quantum adversary can impersonate legitimate Network Functions or User Equipment by forging digital identities and signatures created using classical signing schemes, such as RSA/ECDSA. This can lead to data theft, disruptions in service, and intrusion. Exploiting classical OAuth tokens could also allow an attacker access to protected resources. Furthermore, such an adversary could circumvent mTLS authentication checks and gain access to the service mesh by creating phony X.509 certificates that are seemingly legitimate to other Network Functions (NFs). 

\textbf{Subscriber Impersonation:} Leveraging the capabilities of quantum computers, adversaries can impersonate legitimate network subscribers (User Equipments) by bypassing 5G-AKA mechanisms and decrypting the confidential subscriber identifier (SUCI). 5G-AKA is based on 128-bit symmetric primitives, which are susceptible to brute-force attacks via Grover's algorithm. SUCI, encrypted using the Elliptic Curve Integrated Encryption Scheme (ECIES) with X25519, is vulnerable to quantum attacks.

\textbf{MITM (Man-in-the-Middle) Attack:} Quantum adversaries can place themselves between communicating NFs or between the UE and network infrastructure due to their capability to forge certificates and crack key exchange protocols. This compromises the confidentiality and integrity of communications. Additionally, the attacker can evade detection by traditional security systems by intercepting, decrypting, altering, and re-encrypting in real time---active eavesdropping.

\textbf{Side Channel Attacks:} Some side-channel analysis methods, especially those based on mathematical optimization problems, might be accelerated by quantum computing resources. Even though PQC algorithms present new implementation difficulties, these risks can be mitigated by using proper implementation procedures, such as blinding techniques and constant-time operations.

\subsection{Affected Protocols and Interfaces}

Table~\ref{tab:threats} provides a comprehensive mapping of quantum threats to specific 5G protocols and interfaces, along with the vulnerable cryptographic primitives and recommended post-quantum countermeasures.

\begin{table*}[htbp]
\centering
\caption{Quantum Threats to 5G Protocols and Countermeasures}
\renewcommand{\arraystretch}{1.6}
\setlength{\tabcolsep}{4pt}
\begin{tabular}{>{\raggedright\arraybackslash}p{2.5cm} >{\raggedright\arraybackslash}p{3cm} >{\raggedright\arraybackslash}p{3cm} >{\raggedright\arraybackslash}p{3cm}}
\toprule
\textbf{Protocol} & \textbf{Vulnerable Component} & \textbf{Quantum Threat} & \textbf{PQC Countermeasure} \\
\midrule
\rowcolor[HTML]{D9EAF7}
TLS 1.3 & ECDHE Key Exchange & Shor's Algorithm & ML-KEM (Hybrid) \\
TLS 1.3 & ECDSA/RSA Signatures & Shor's Algorithm & ML-DSA \\
\rowcolor[HTML]{D9EAF7}
DTLS & ECDHE Key Exchange & Shor's Algorithm & ML-KEM (Hybrid) \\
DTLS & ECDSA/RSA Signatures & Shor's Algorithm & ML-DSA \\
\rowcolor[HTML]{D9EAF7}
mTLS & X.509 Certificates & Signature Forgery & ML-DSA Certificates \\
mTLS & Key Exchange & MITM Attack & ML-KEM (Hybrid) \\
\rowcolor[HTML]{D9EAF7}
IKEv2 & DH/ECDH Groups & Key Compromise & ML-KEM + PPK \\
IKEv2 & Certificate Auth & Impersonation & ML-DSA Certificates \\
\rowcolor[HTML]{D9EAF7}
OAuth 2.0 & JWT Signatures & Unauthorized Access & ML-DSA-JWS \\
ECIES (SUCI) & ECC Encryption & Identity Exposure & ML-KEM SUCI \\
\bottomrule
\end{tabular}
\label{tab:threats}
\end{table*}

\textbf{TLS/DTLS:} Both protocols use ECDHE for key exchange and ECDSA/RSA for certificate signatures. Quantum attacks can break both primitives, compromising session keys and enabling certificate forgery. This affects all SBI communications (mTLS over TLS) and N2 control plane messages (DTLS).

\textbf{IKEv2 (IPsec):} Used to secure N2 and N3 interfaces between gNB and 5GC (AMF/UPF), IKEv2 employs Diffie-Hellman groups for key establishment. Quantum computers can solve the discrete logarithm problem, breaking these groups and exposing IPsec tunnel keys, compromising both control plane and user plane data.

\textbf{OAuth 2.0:} JSON Web Tokens (JWT) used for authorization are signed with RSA or ECDSA. Quantum signature forgery enables attackers to create fraudulent authorization tokens, gaining unauthorized access to network services and resources.

\textbf{mTLS:} The dual authentication in mTLS relies on bidirectional certificate verification using classical signatures. Quantum attacks enable bilateral impersonation, completely defeating the mutual authentication mechanism.

The preceding threat analysis clearly demonstrates that quantum computing poses a fundamental risk to the 5G security architecture, thereby mandating a proactive migration to post-quantum cryptographic primitives across all critical protocol layers and network interfaces.

\section{Proposed Framework}
The proposed QORE framework is specifically focused on upgrading traditional security protocols to achieve quantum resistance, beginning with the SBI communication framework. All essential security protocols should have a thorough PQ implementation to guarantee backward compatibility and uninterrupted operations during the transition. Using Quantum Random Number Generators (QRNGs), which are in charge of cryptographic seed generation in the suggested protocols, we present an approach that ensures a true entropy source and improves the overall security and unpredictability of the key materials that will be generated and used in the communications.

\subsection{PQ-TLS: Post-Quantum Transport Layer Security}

Post-Quantum TLS 1.3 is an extended version of the classical TLS 1.3 protocol (discussed above), which guarantees security against either active or passive quantum adversaries. It achieves this by integrating a set of PQ Primitives, such as ML-KEM and ML-DSA, in the key agreement and signing messages, including X.509 certificates. The upgradation to post-quantum primitives retains compatibility with TLS 1.3 standards, thereby providing a seamless integration with existing networks.

\textbf{Architecture:} Despite reusing the core functionality and handshake sequence of the classical TLS messages, PQ-TLS 1.3 still achieves quantum-resistance by strategically replacing quantum-vulnerable primitives with their post-quantum counterparts. This is accomplished by modifying \texttt{ClientHello}---to negotiate and share ephemeral hybrid PQ keys, \texttt{ServerHello}---to send the encapsulated ciphertext, and share a PQ signed X.509 Server Certificate.

\textbf{Certificate Infrastructure:} We focus on implementing a hybrid certificate strategy capable of supporting both homogeneous (pure post-quantum) and hybrid (post-quantum + classical) signing algorithms and subject public keys.

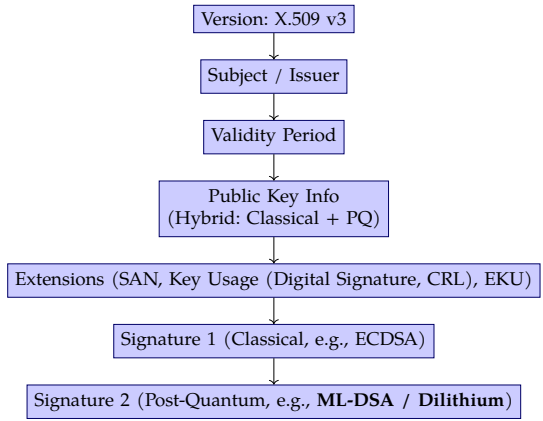
\begin{figure}[h]
\centering
\scalebox{0.7}{
\begin{tikzpicture}[node distance=0.5cm, every node/.style={draw=blue!40!black, fill=blue!20, rectangle, align=center}, arrow/.style={-Stealth, thick}, inner sep=4pt]

\node (version) {Version: X.509 v3};
\node (subject) [below=of version] {Subject / Issuer};
\node (validity) [below=of subject] {Validity Period};
\node (pubkey) [below=of validity] {Public Key Info\\(Hybrid: Classical + PQ)};
\node (extensions) [below=of pubkey] {Extensions (SAN, Key Usage (Digital Signature, CRL), EKU)};
\node (sig1) [below=of extensions] {Signature 1 (Classical, e.g., ECDSA)};
\node (sig2) [below=of sig1] {Signature 2 (Post-Quantum, e.g., \textbf{ML-DSA / Dilithium})};
\draw[->] (version) -- (subject);
\draw[->] (subject) -- (validity);
\draw[->] (validity) -- (pubkey);
\draw[->] (pubkey) -- (extensions);
\draw[->] (extensions) -- (sig1);
\draw[->] (sig1) -- (sig2);


\end{tikzpicture}
}
\caption{Structure of a hybrid post-quantum X.509 certificate with classical and post-quantum signatures.}
\label{fig:hybrid-x509}
\end{figure}

Certificate Signing Requests (CSRs) comprising ML-DSA-65 Keys as their verification keys and other details of the end-entity are signed by the Root or the Intermediate CA using Post-Quantum schemes, such as ML-DSA-65/87, or SLH\_DSA\_PURE\_SHA2\_192S, thereby offering estimated security levels equivalent to AES-192 and AES-256, respectively. Furthermore, following the guidance from RFC drafts concerning post-quantum X.509 certificates \cite{draft-ietf-lamps-dilithium-certificates}, ML-DSA keys are meticulously embedded as public key algorithm identifiers within standard X.509 certificate structures, ensuring immediate compatibility with existing Public Key Infrastructure (PKI) systems. Certificates include standard extensions such as Subject Alternative Name (SAN), Key Usage, and Extended Key Usage, and can be validated by existing TLS libraries after extending the trust model to recognize post-quantum algorithm OIDs. Figure~\ref{fig:hybrid-x509} demonstrates a standard Hybrid PQ X.509 Certificate.

\begin{figure}[h]
\centering
\scalebox{0.6}{
\begin{tikzpicture}[every node/.style={draw=blue!40!black, fill=blue!15, rectangle, rounded corners, align=left}, node distance=0.7cm]

\node (priv) {Private Key (2464 bytes)\\
ML-KEM-768: 2400 B, X25519 Priv: 32 B, X25519 Pub: 32 B};

\node (pub) [below=of priv] {Public Key (1216 bytes)\\
ML-KEM-768: 1184 B, X25519: 32 B};

\node (ct) [below=of pub] {Ciphertext (1120 bytes)\\
ML-KEM-768: 1088 B, X25519: 32 B};

\node (shared) [below=of ct, fill=blue!25] {Shared Key (32 bytes)\\
HKDF.Extract(salt, secret:ML-KEM-768 shared + X25519 shared)};

\draw[->, thick] (priv) -- (pub);
\draw[->, thick] (pub) -- (ct);
\draw[->, thick] (ct) -- (shared);

\end{tikzpicture}
}
\caption{Overview of X25519MLKEM768 hybrid key exchange: private, public, ciphertext, and shared secret contributions.}
\label{fig:hybrid-pq-single-side}
\end{figure}
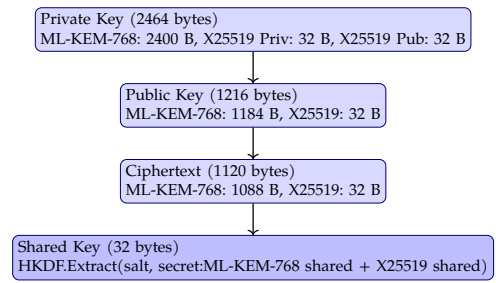

\textbf{Key Exchange Mechanism:} We implement hybrid ML-KEM key encapsulation combined with X25519 -- \textbf{X25519MLKEM768} \cite{draft-ietf-tls-ecdhe-mlkem}  to provide defense-in-depth. The hybrid approach ensures security if either primitive remains secure, mitigating risks from potential cryptographic breaks or implementation vulnerabilities. The key exchange generates a shared secret combining contributions from both algorithms, which feeds into HKDF \cite{rfc5869} for deriving encryption and MAC keys. A flow diagram is shown in Figure~\ref{fig:hybrid-pq-single-side}.

\textbf{Cipher Suite Configuration:} Our PQ-TLS supports multiple cipher suites balancing security and performance:
\begin{itemize}
\item TLS\_AES\_256\_GCM\_SHA384 
\item TLS\_AES\_128\_GCM\_SHA256 
\item TLS\_CHACHA20\_POLY1305\_SHA256 
\end{itemize}


\begin{table*}[htbp]
\centering
\caption{Comparison of Classical and Post-Quantum TLS Mechanisms}
\renewcommand{\arraystretch}{1.25}
\setlength{\tabcolsep}{5pt}
\rowcolors{2}{blue!5}{white}
\begin{tabular}{>{\raggedright\arraybackslash}p{3.2cm} >{\raggedright\arraybackslash}p{5.5cm} >{\raggedright\arraybackslash}p{5.5cm}}
\toprule
\rowcolor{blue!15}
\textbf{Field} & \textbf{Classical Configuration} & \textbf{Post-Quantum Configuration} \\
\midrule

CA Type & Private (Internal) CA & Private (Internal) CA \\

Certificate Signature Algorithm & RSA/ECDSA classical signature algorithms & \textbf{Homogeneous: ML-DSA-44/65/87}\\
& & Hybrid: ML-DSA-Ed448, ML-DSA-Ed25519 + Any TLS v1.3/1.2 classical signature schemes \\

Signature Length & 64 octets (Ed25519 example) & 3293 octets + Classical Signature Length (e.g., 64 octets for Ed25519) \\

Key Exchange Mechanism & ECDHE / DH key exchange & \textbf{Homogeneous: ML-KEM (512/768/1024)}\\
& & Hybrid: ECDHE\_ML-KEM (e.g., X25519MLKEM768) \\

Key Exchange Length (Public key) & 32 octets (X25519/P256) & 1216 octets (ML-KEM-768: 1184 octets, X25519/P256: 32 octets) \\

Key Exchange Ciphertext Length & 32 octets & 1120 Octets (ML-KEM-768: 1088 Octets, X25519/P256: 32 Octets)  \\

Hybrid KEM KDF algorithm & N/A & HKDF (SHAKE256/SHA-3) \\

AEAD - Symmetric Encryption \& Authentication Algorithm & AES256\_GCM, ChaCha20\_Poly1305 & AES256\_GCM, ChaCha20\_Poly1305 \\

enckeylen & 32 octets (256 bits) & 32 octets (256 bits) \\

ivlen & 12 octets (96 bits) & 12 octets (96 bits) \\

mackeylen & 32/48 octets (256/384 bits) & 32/48 octets (256/384 bits) \\

maclen & 32/48 octets (256/384 bits) & 32/48 octets (256/384 bits) \\

TLS Key Derivation Function (KDF) & HMAC-based HKDF / PRF (TLS v1.2) & HMAC-based Expand \& Extract KDF (HKDF) / PRF for TLS v1.2 \\

TLS Finished MAC Algorithm & HMAC-SHA-256/384 & HMAC-SHA-256/384 \\

Fallback Methods & TLS v1.2, Classical signature \& key exchange schemes supported & PQ-aware fallback with hybrid algorithms supported \\

\bottomrule
\end{tabular}
\label{tab:pqtls}
\end{table*}


\textbf{Symmetric Cryptography:} We retain AES-256-GCM and ChaCha20-Poly1305 for authenticated encryption as these symmetric algorithms provide adequate quantum resistance through Grover's algorithm considerations (requiring 256-bit keys for 128-bit post-quantum security) \cite{nist_grover_aes} \cite{ramos-calderer2021}.

\begin{figure*}
    \centering
    \includegraphics[width=1.0\linewidth]{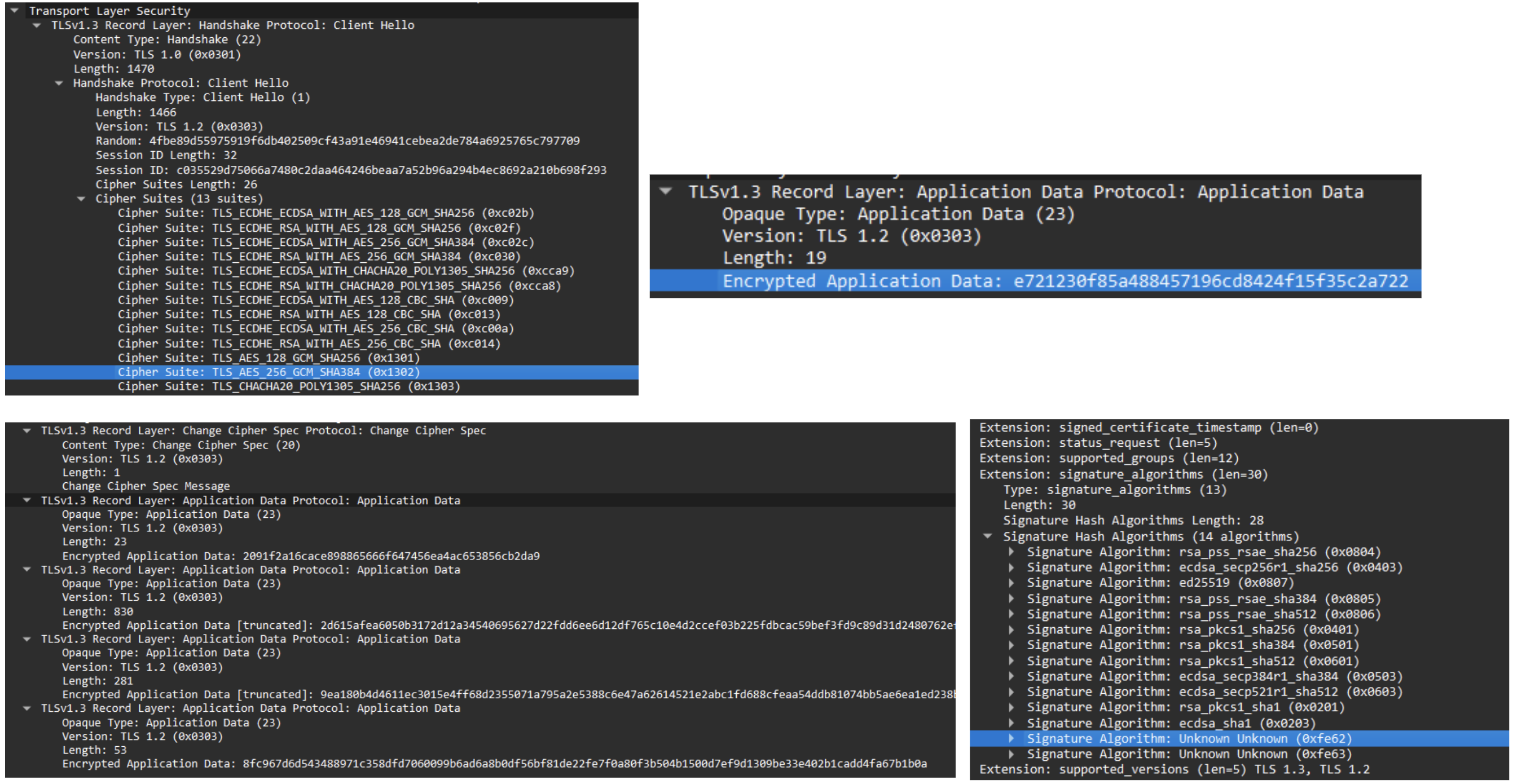}
    \caption{PQ-TLS 1.3 PCAP}
    \label{fig:tls_pcap}
\end{figure*}

\begin{figure*}
    \centering
    \includegraphics[width=1.0\linewidth]{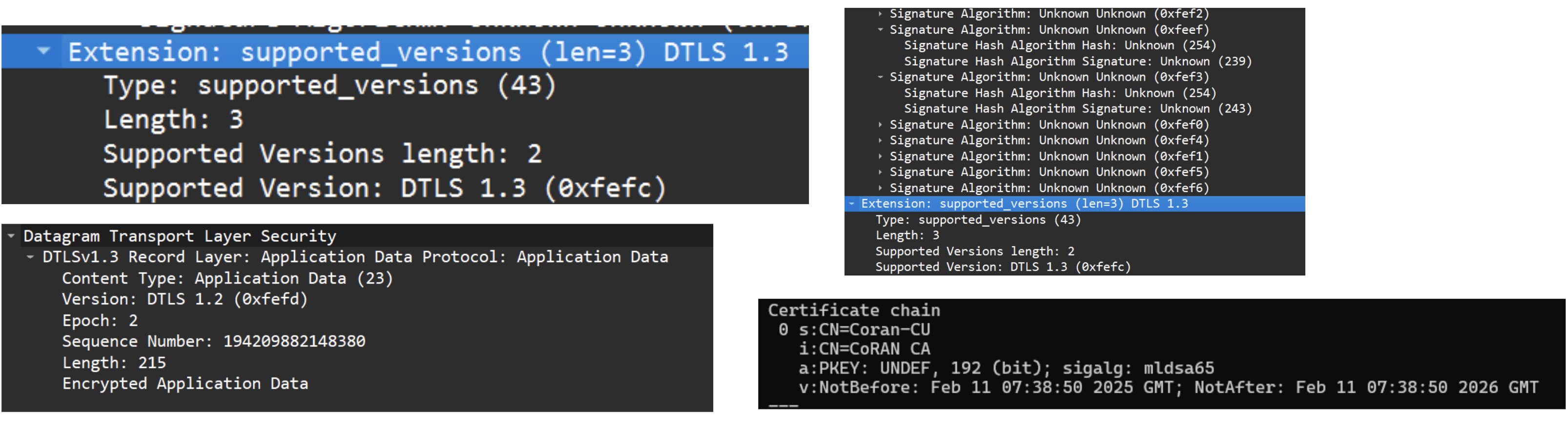}
    \caption{PQ-DTLS PCAP}
    \label{fig:dtls_pcap}
\end{figure*}

Table~\ref{tab:pqtls} presents a comprehensive comparison between classical and post-quantum TLS configurations, highlighting changes in certificate signatures, key exchange mechanisms, and resultant message sizes.\\
\textbf{\textbf{Packet Captures for PQ-TLS 1.3 and PQ-DTLS 1.3 are depicted in Figure~\ref{fig:tls_pcap} and Figure~\ref{fig:dtls_pcap}}
}

\subsection{PQ-IPsec: Post-Quantum Internet Protocol Security}

IPsec secures the critical N2 (gNB to AMF) and N3 (gNB to UPF) interfaces in 5G networks, protecting both control plane signaling and user plane data transmission. Our PQ-IPsec implementation upgrades the IKEv2 protocol with quantum-resistant key exchange while maintaining tunnel security and performance.

\begin{figure*}
    \centering
    \scalebox{0.8}{
    \begin{tikzpicture}[
  node distance=1.4cm,
  every node/.style={
    draw=black!30,
    rounded corners,
    align=center,
    text width=6cm,
    minimum height=1.2cm,
    inner sep=6pt,
    font=\small,
  },
  process/.style={fill=green!15, draw=green!40!black},
  startstop/.style={fill=blue!15, draw=blue!40!black, text width=4cm},
  note/.style={draw=black!50, fill=yellow!10, text width=10cm, align=left, font=\footnotesize, rounded corners},
  arrow/.style={-Stealth, thick}
]

\node[startstop] (client) {Initiator \\ \textit{(e.g., gNB)}};
\node[startstop, right=6cm of client] (server) {Responder \\ \textit{(e.g., AMF/UPF)}};

\draw[arrow] (client) -- (server) node[midway, above, yshift=2pt, font=\scriptsize, align=center] {
HDR, SAi (IKE proposals: AES-CBC-256, HMAC-SHA2-256),\\
KEi, Ni $\rightarrow$ (IKE/Child SA proposal)\\
\textbf{N2/N3}
};

\draw[arrow, dashed] (server) -- (client) node[midway, below, yshift=-2pt, font=\scriptsize, align=center] {
HDR, SAr, KEr, Nr, [CERTREQ],
};

\def\ysep{0.5cm}

{\fontfamily{qpl}\selectfont
This text uses a different font typeface

\node[process, below=1.8cm of $(client)!0.5!(server)$] (ikeinit) {%
\textbf{IKE\_SA\_INIT}\\[3pt]
Establishes initial security association and negotiates cryptographic parameters.
};

\node[process, below=\ysep of ikeinit] (intermediate) {%
\textbf{IKE\_INTERMEDIATE}\\[3pt]
\textbf{(RFC 9370: ML-KEM-768)} — Performs post-quantum key exchange using ML-KEM.
};

\node[process, below=\ysep of intermediate] (auth) {%
\textbf{IKE\_AUTH}\\[3pt]
\textbf{(RFC 8784: PQ Pre-Shared Keys)} — Authenticates peers using Post-Quantum pre-shared keys (PPK).
};

\node[process, below=\ysep of auth] (childsa) {%
\textbf{CREATE\_CHILD\_SA}\\[3pt]
Used to rekey or create additional child security associations.
};

\node[process, below=\ysep of childsa] (inform) {%
\textbf{INFORMATIONAL}\\[3pt]
Used for notifications, deletion, and NAT detection.
};

\node[note, below=0.8cm of inform] (notes) {
\textbf{Notes:}\\[2pt]
– IKE SA = Encrypted control channel for IKE messages.\\
– Child SAs = ESP/AH tunnels carrying application data.\\
– Proposals in IKE\_SA\_INIT are for IKE SA.\\
– Proposals in IKE\_AUTH or CREATE\_CHILD\_SA are for ESP/AH (IPsec).\\
– \textbf{ML-KEM-768} provides PQ key exchange (RFC 9370).\\
– \textbf{PPK} adds PQ pre-shared authentication (RFC 8784).
};

\node[draw=black!50, rounded corners, inner sep=27.2pt, fit=(client) (server) (notes) (ikeinit) (intermediate)  (auth) (childsa) (inform)] {};
}

\foreach \n in {ikeinit, intermediate, auth, childsa, inform} {
    \draw[<-, very thick, black] 
        ($(\n.west)+(-1.0cm,0)$) -- ($(\n.west)+(-0.2cm,0)$);
    
    \draw[->, very thick, black] 
        ($(\n.east)+(0.2cm,0)$) -- ($(\n.east)+(1.0cm,0)$);
}

\end{tikzpicture}
}
\vspace{0.35cm} 

 \includegraphics[width=\textwidth]{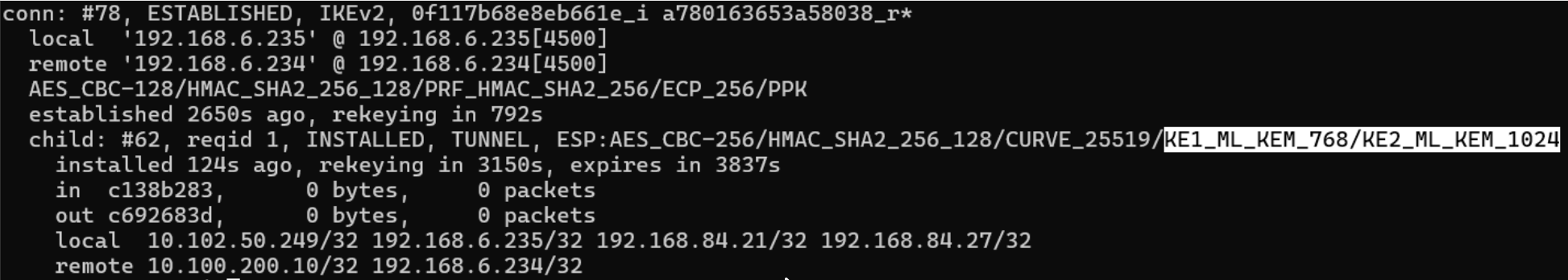}

  \caption{PQ-IPsec Message Flow. The diagram above illustrates the IKEv2 handshake with PQ Pre-Shared Keys (PPK) and hybrid key exchange. The flow shows: \\(1) IKE\_SA\_INIT, which uses classical DH to negotiate the parameters for subsequent messages,\\(2) \textbf{IKE\_INTERMEDIATE}, which carries out the ML-KEM-768 Key Exchange to form a shared secret that will be mixed in the final IKE secret,\\(3) IKE\_AUTH exchange to perform authentication and certificate verification using PPKs and ML-DSA respectively, \\(3) CREATE\_CHILD\_SA for IPsec SA establishment, and\\ (4) an encrypted ESP tunnel for data transmission. This architecture is catered to support multiple concurrent key exchanges, thereby meeting the throughput demands of 5G.}
    \label{fig:pq-ipsec}
\end{figure*}


\begin{table*}[htbp]
\centering
\caption{Comparison of PQ-IPsec Shared Key (PPK) and Classical Mechanisms}
\renewcommand{\arraystretch}{1.25}
\setlength{\tabcolsep}{5pt}
\rowcolors{2}{blue!5}{white}
\begin{tabular}{>{\raggedright\arraybackslash}p{3.2cm} >{\raggedright\arraybackslash}p{5.5cm} >{\raggedright\arraybackslash}p{5.5cm}}
\toprule
\rowcolor{blue!15}
\textbf{Field} & \textbf{Classical Configuration} & \textbf{Post-Quantum Configuration (PPK)} \\
\midrule

CA Type & Private (Internal) CA & Private (Internal) CA \\

Authentication & Classical PKI certificates (x.509) & \textbf{PQ-Pre Shared Keys (PPKs) – 256 bits} \\

Key Exchange Mechanism & DH-3072 / ECP-384 / X25519 & DH-3072 / ECP-384 / X25519 + ML-KEM-768 (up to 7 key exchanges) \\

Key Exchange Length (Public key) & 32/64 octets & 1216 octets+ (same as ML-KEM hybrid mode) \\

Key Exchange CipherText Length & 32 octets & 1120 octets \\

AEAD - Symmetric Encryption \& Authentication Algorithm & AES256-GCM, AES-CCM, ChaCha20-Poly1305 & AES256-GCM, AES-CCM, ChaCha20-Poly1305 \\

enckeylen & 32 octets (256 bits) & 32 octets (256 bits) \\

ivlen & 12 octets (96 bits) & 12 octets (96 bits) \\

mackeylen & 32/48 octets (256/384 bits) & 32/48 octets (256/384 bits) \\

maclen & 32/48 octets (256/384 bits) & 32/48 octets (256/384 bits) \\

Key Derivation Function & HMAC-SHA-384 & HMAC-SHA-384/SHAKE256 \\

Random Number Generator & CSPRNG, RNG, DRBG-HMAC-SHA-256 & QRNG, TRNG\\

\bottomrule
\end{tabular}
\label{tab:ppk}
\end{table*}

\textbf{Post-Quantum Pre-Shared Keys (PPK):} We introduce PPK as the primary authentication mechanism for PQ-IPsec, as detailed in Table~\ref{tab:ppk}. PPKs are 256-bit quantum-resistant shared secrets distributed through secure out-of-band channels during network function provisioning. This approach provides immediate quantum resistance without requiring complex PKI migrations for point-to-point connections.

\textbf{Hybrid Key Exchange:} Our implementation supports up to 7 key exchange methods simultaneously, combining classical and post-quantum algorithms:
\begin{itemize}
\item ML-KEM-768 (Primary PQ algorithm)
\item X25519 or ECP-384 (Classical fallback)
\item DH-2048 (Legacy support)
\end{itemize}

IKEv2 combines key-exchange outputs by updating the SK\_* values with the KDF after each exchange; post-quantum pre-shared keys (PPKs) are mixed into SK\_d, SK\_pi, and SK\_pr (i.e., into the PRF/KDF inputs) per RFC 8784, ensuring the derived keying material depends on the PPK as well as the (EC)DH/PQC contributions. \cite{rfc9370,rfc8784}

\textbf{Child SA Security:} IPsec tunnel encryption uses AES-256-GCM with 256-bit keys, providing adequate quantum resistance for user and control plane data. The key derivation incorporates PPK material, ensuring quantum security even if the initial key exchange is compromised.

\begin{figure*}
    \centering
    \includegraphics[width=1.0\textwidth]{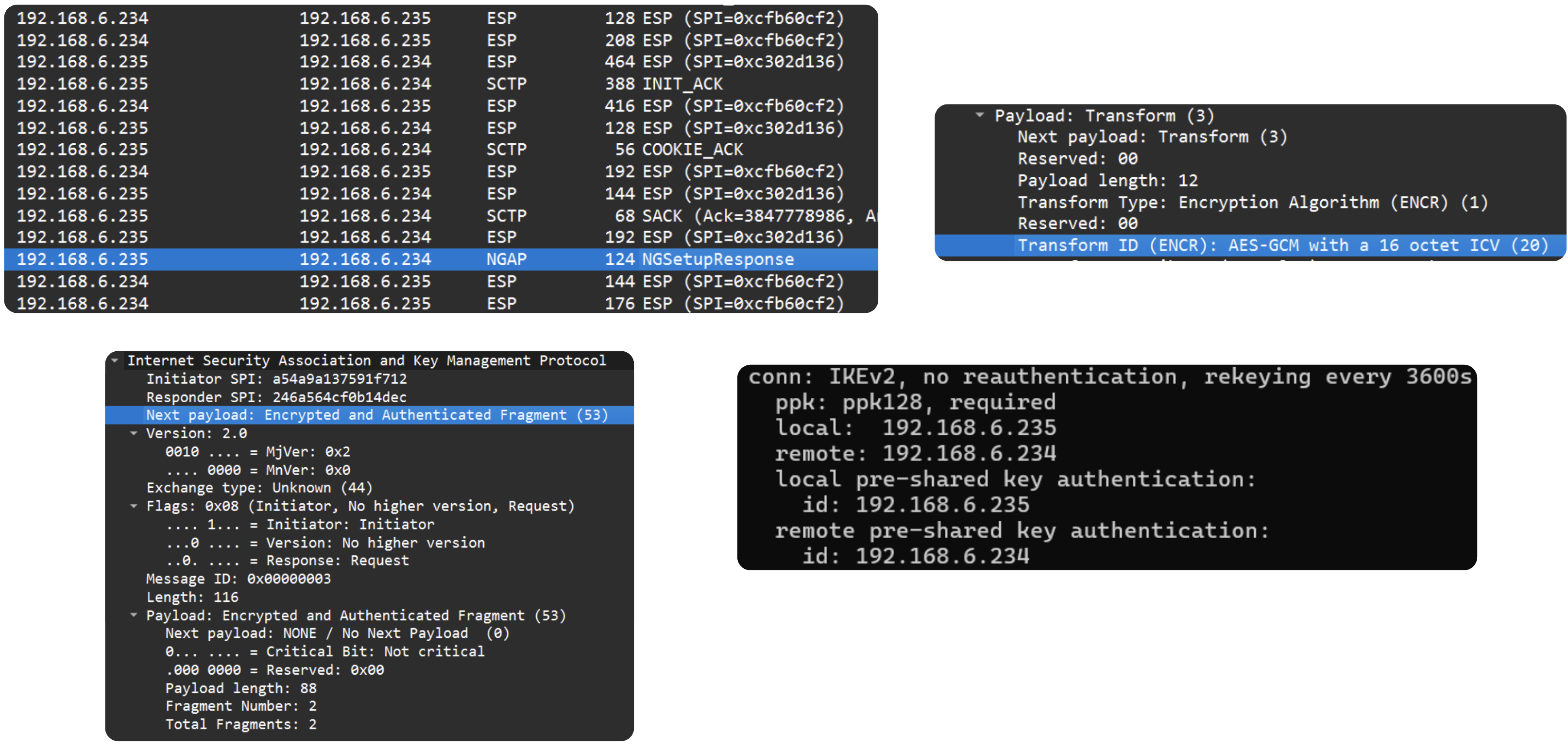}
    \caption{Packet capture showing IPsec (PQ) and NGAP packets}
    \label{fig:ipsec_pcap}
\end{figure*}

\textbf{Performance Optimization:} Figure~\ref{fig:pq-ipsec} demonstrates efficient SA rekey mechanisms supporting high-throughput N3 user plane traffic. The implementation pre-negotiates Child SAs before expiration, ensuring seamless transitions without packet loss.\\
\textbf{Packet Captures and XFRM policies for PQ-IPsec can be seen in Figure~\ref{fig:ipsec_pcap}}

\subsection{PQ-mTLS: Post-Quantum Mutual Transport Layer Security}

Mutual TLS forms the security foundation for Service-Based Interface communications between Network Functions. Our PQ-mTLS implementation extends PQ-TLS with bilateral authentication requirements and NF-specific security policies.

\begin{figure*}
    \centering
    \includegraphics[width=0.95\textwidth]{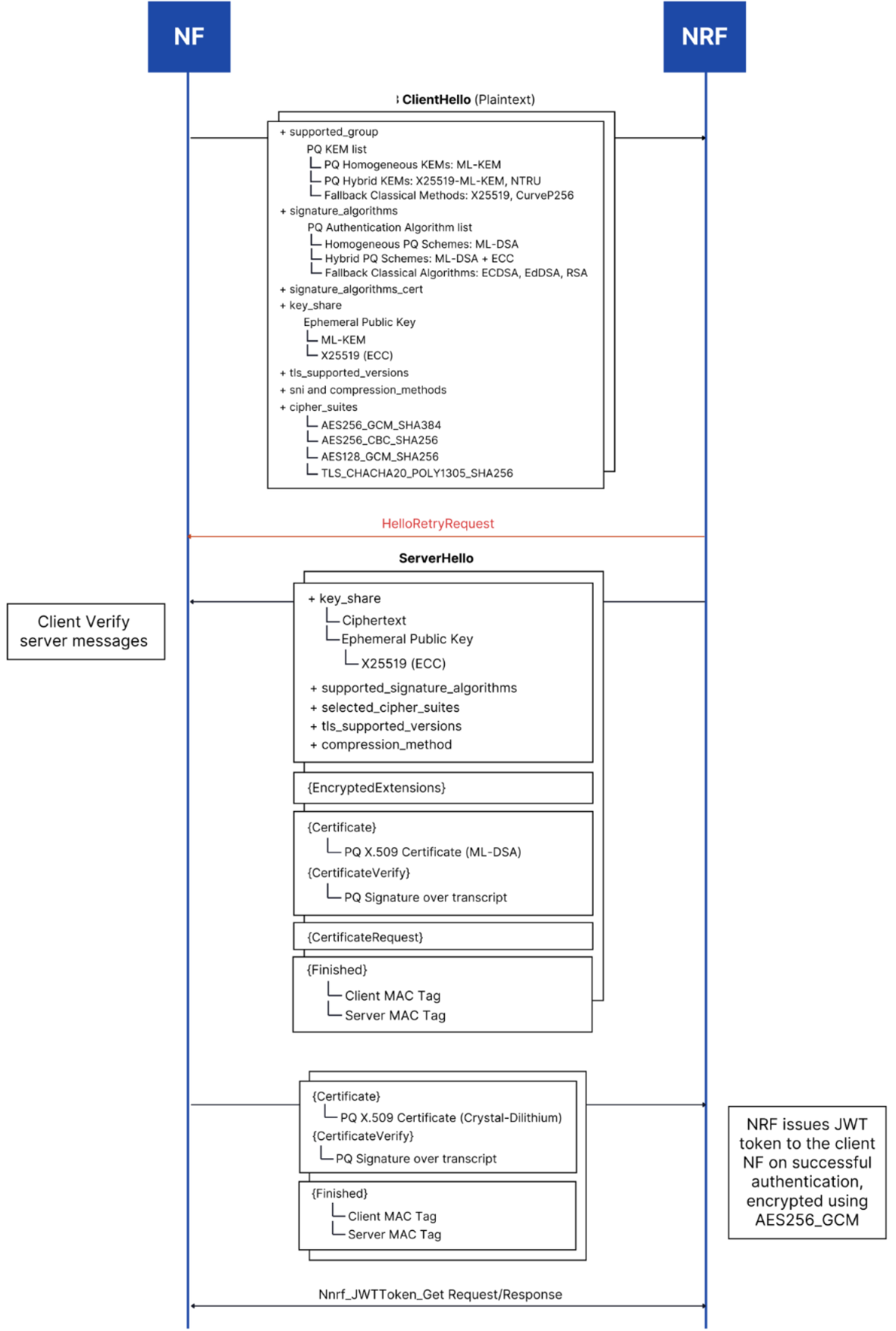}
    \caption{A sequence of layered steps is followed by the flow: 1)First, ML-DSA-signed X.509 certificates are used for mutual authentication between the consumer and producer NFs via PQ-mTLS. 2)Secondly, the consumer NF obtains an ML-DSA-signed OAuth token from the NRF with the relevant policies and scopes.3)The ML-DSA signed OAuth token is presented by the Consumer for application-layer authorization; (4) the Producer NF verifies the token signature using the NRF's OAuth2.0 public key (available at JWKS endpoint \cite{rfc7517}) and provides access; (5) secure communication then takes place over the PQ-TLS encrypted channel. This token is certificate-bound, as described in \cite{rfc8705}, and is transported via the secure transport of mTLS.}
    \label{fig:pq-mtls-oauth}
\end{figure*}

\textbf{Mutual Certificate Authentication:} Both client and server NFs present ML-DSA signed certificates during the handshake. Each NF validates the peer's certificate chain, verifies ML-DSA signatures, checks certificate revocation status via OCSP or CRL (using ML-DSA signed responses), and confirms subject identity matches the expected NF identifier. The CRLs and OCSP are published and digitally signed by an issuing CA, which may be delegated by the Root CA, allowing clients to verify certificate revocation status without contacting the Root CA directly.

\textbf{Certificate Lifecycle Management:}  Certificates have shorter validity periods (3-6 months) to reduce key exposure and enforce regular key rotation. Automated certificate management using ACME protocol (adapted for ML-DSA signatures) ensures seamless certificate rotation, revocation, and issuance without service disruption.

\textbf{Zero-Trust Architecture:} PQ-mTLS implements a zero-trust security model \cite{nist_sp800_207} where every inter-NF communication requires mutual authentication and authorization. Combined with PQ-OAuth (section IV-D), this provides comprehensive access control preventing lateral movement within the 5GC even if individual NFs are compromised.

\textbf{NF-Specific Policies:} Different NF types (AMF, SMF, UPF, etc.) have tailored security policies enforced during the mTLS handshake. For example, UPF to UPF connections may use different cipher suites optimized for high-throughput data plane traffic compared to AMF to AUSF control plane connections. The post-quantum versions of these ciphersuites can be AEGIS \cite{draft-denis-tls-aegis, draft-irtf-cfrg-aegis-aead}, ASCON \cite{nist_ascon}, etc.

\subsection{PQ-OAUTH: Post-Quantum OAuth 2.0}

OAuth 2.0 provides an authorization framework for service-based communications in 5GC. Our PQ-OAuth implementation migrates JWT token signatures to ML-DSA while maintaining compatibility with existing OAuth workflows.

\begin{figure*}[htbp]
    \centering
    \includegraphics[width=0.7\textwidth,height=15cm]{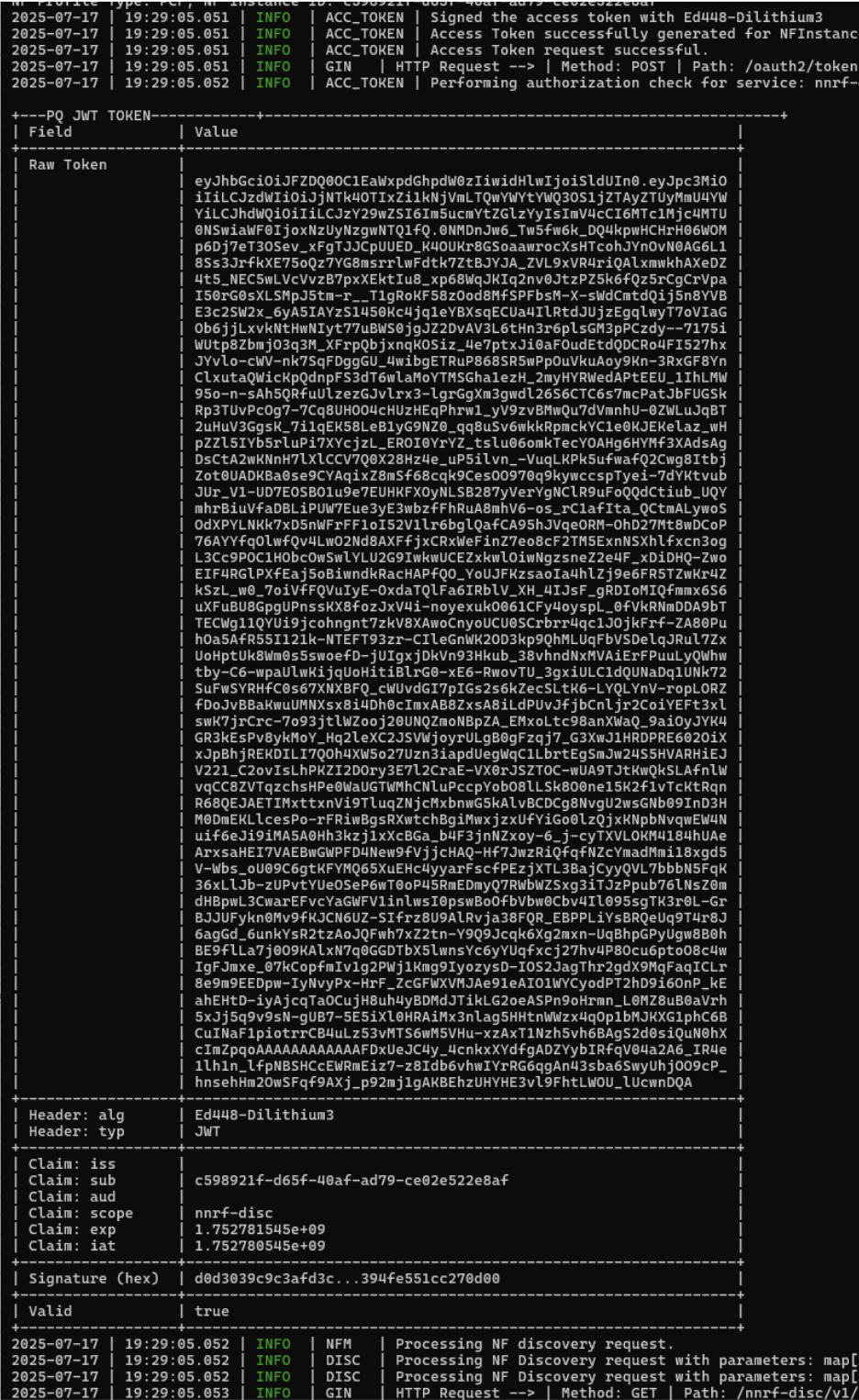}
    \caption{PQ-OAuth token creation and signing. The authorization server, typically NRF, generates OAuth access tokens as JSON Web Tokens (JWT). The signature algorithm, ML-DSA-65, is specified in the token header. Standard claims like subject, audience, expiration, and scope as well as unique 5G specific claims like NF instance ID, permitted services, and resource permissions are included in the payload. NRF's ML-DSA private key is used to sign the token,through which a PQ secure signature is produced. Before allowing access the authenticity of the NRF with its public key and consumer NFs show this token to access producer NF services.}
    \label{fig:pq-oauth}
\end{figure*}

As illustrated in Figure~\ref{fig:pq-oauth}, the PQ-OAuth token generation process strategically extends the standard OAuth 2.0 authorization flow to securely incorporate PQ digital signature algorithms. The Network Repository Function (NRF), acting as the authorization server, issues JSON Web Tokens (JWTs) signed with ML-DSA, ensuring resilience against quantum attacks.

\begin{figure*}[htbp]
    \centering
    \includegraphics[width=0.5\linewidth]{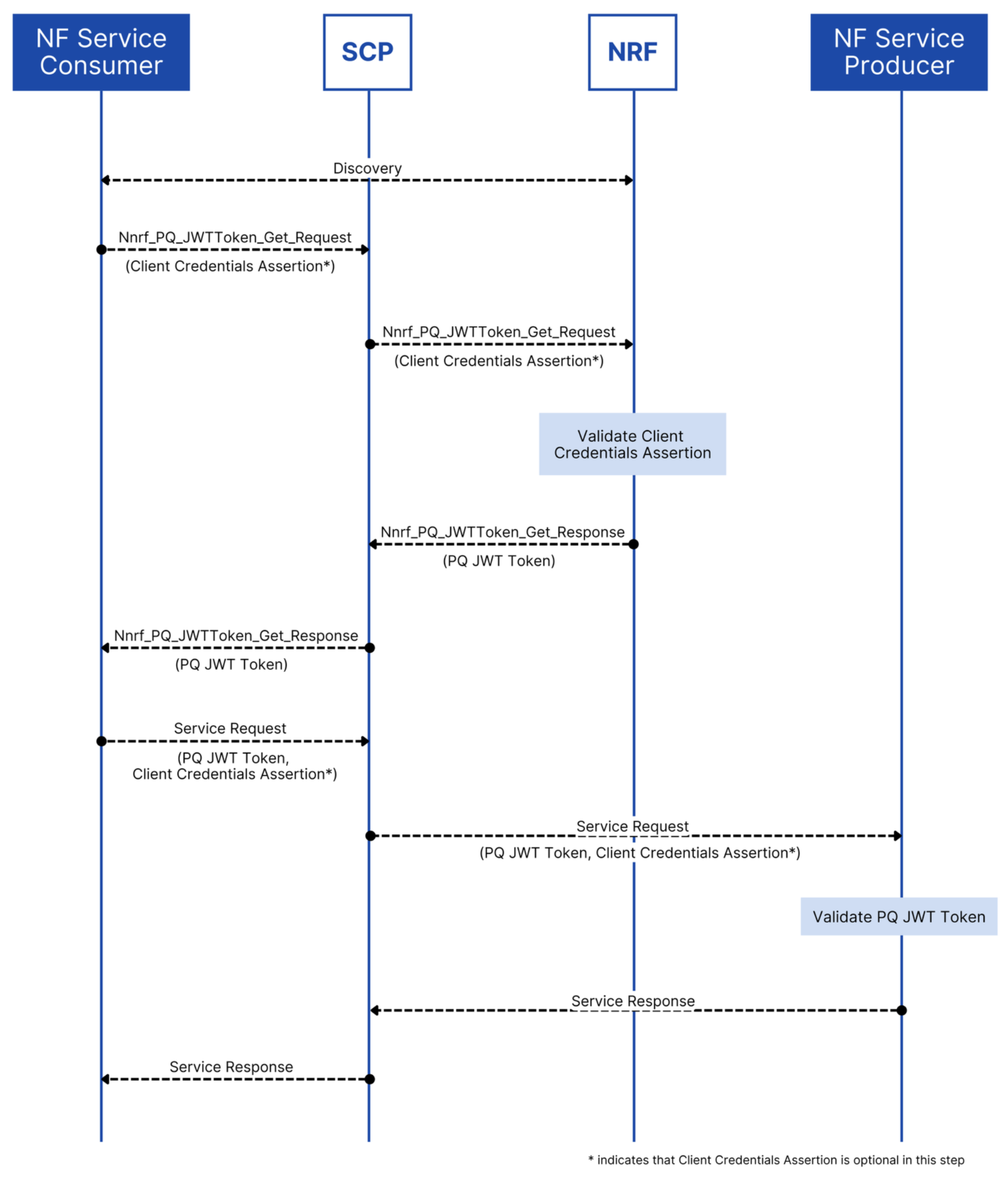}
    \caption{PQ-OAuth Token Acquisition and Validation Procedure. This sequence diagram illustrates the process for obtaining and using a Post-Quantum (PQ) secured OAuth JSON Web Token (JWT) between 5G Network Functions (NFs). The NF Service Consumer initiates a Nnrf\_PQ\_JWTToken\_Get\_Request (optionally including a Client Credentials Assertion) via the Service Communication Proxy (SCP) to the Network Repository Function (NRF). The NRF validates the client credentials, then issues a PQ-signed JWT (Nnrf\_PQ\_JWTToken\_Get\_Response) using a post-quantum signature algorithm (e.g., ML-DSA). The consumer includes this PQ JWT in subsequent service requests to the NF Service Producer, either directly or via the SCP. Upon receipt, the producer validates the PQ JWT using the NRF’s public key to confirm authenticity and authorization before processing the service request. The described flow ensures end-to-end, post-quantum-secure authentication and authorization between 5G network functions.}
    \label{fig:oauth flow}
\end{figure*}

The overall acquisition and validation procedure for PQ-OAuth tokens is shown in Figure~\ref{fig:oauth flow}. In this flow, Network Functions (NFs) obtain PQ-signed JWTs from the NRF via the Service Communication Proxy (SCP) \cite{3gpp_33522, 3gpp_23501} and present them to other NFs for service access. The receiving NF validates the ML-DSA signature using the NRF’s public key before granting access, providing end-to-end post-quantum-secure authorization. 

Following 3GPP TS 33.522 \cite{3gpp_33522}, the procedure ensures that tokens issued by the NRF are securely verified by each NF. By leveraging the SCP and post-quantum signatures, the framework maintains secure service access and prevents unauthorized interactions between NFs, even in a quantum-threat environment. A demonstration of the implemented architecture is publicly available at \textbf{\href{https://www.youtube.com/watch?v=JM4s2G23w5I}{YouTube}}.

\textbf{Token Structure:} PQ-OAuth tokens maintain the three-part JWT structure (header.payload.signature) but use ML-DSA for signature generation. The header specifies algorithm "ML-DSA-65" or "ML-DSA-87", the payload contains standard OAuth claims plus 5G-specific attributes (NF type, allowed services, instance ID), and the signature is computed over the concatenated header and payload using the authorization server's ML-DSA private key.

\textbf{Token Lifecycle:} Tokens have short validity periods (15–60 minutes) to limit exposure windows. The authorization server (NRF) issues tokens upon NF authentication and can revoke tokens by publishing revocation lists signed with ML-DSA. Consumer NFs cache tokens for efficiency but must refresh before expiration.

\textbf{Token Validation:} Producer NFs validate incoming tokens by: (1) parsing the JWT structure, (2) extracting the ML-DSA public key from the authorization server (cached with periodic refresh), (3) verifying the ML-DSA signature over header.payload, (4) checking token expiration and revocation status, and (5) validating claims against request context (audience, scope, subject).

\textbf{Integration with mTLS:} As demonstrated in Figure~\ref{fig:pq-mtls-oauth}, PQ-OAuth and PQ-mTLS operate in concert to provide comprehensive security. In particular, OAuth offers fine-grained authorization at the application layer, whereas mTLS is in charge of authenticating both communicating parties at the transport layer. Defense-in-depth is successfully ensured by this potent, tiered strategy against both enduring classical vulnerabilities (such as token theft and malicious endpoints) and new quantum threats.

\begin{figure*}[!t]
    \centering
    \includegraphics[width=0.9\linewidth]{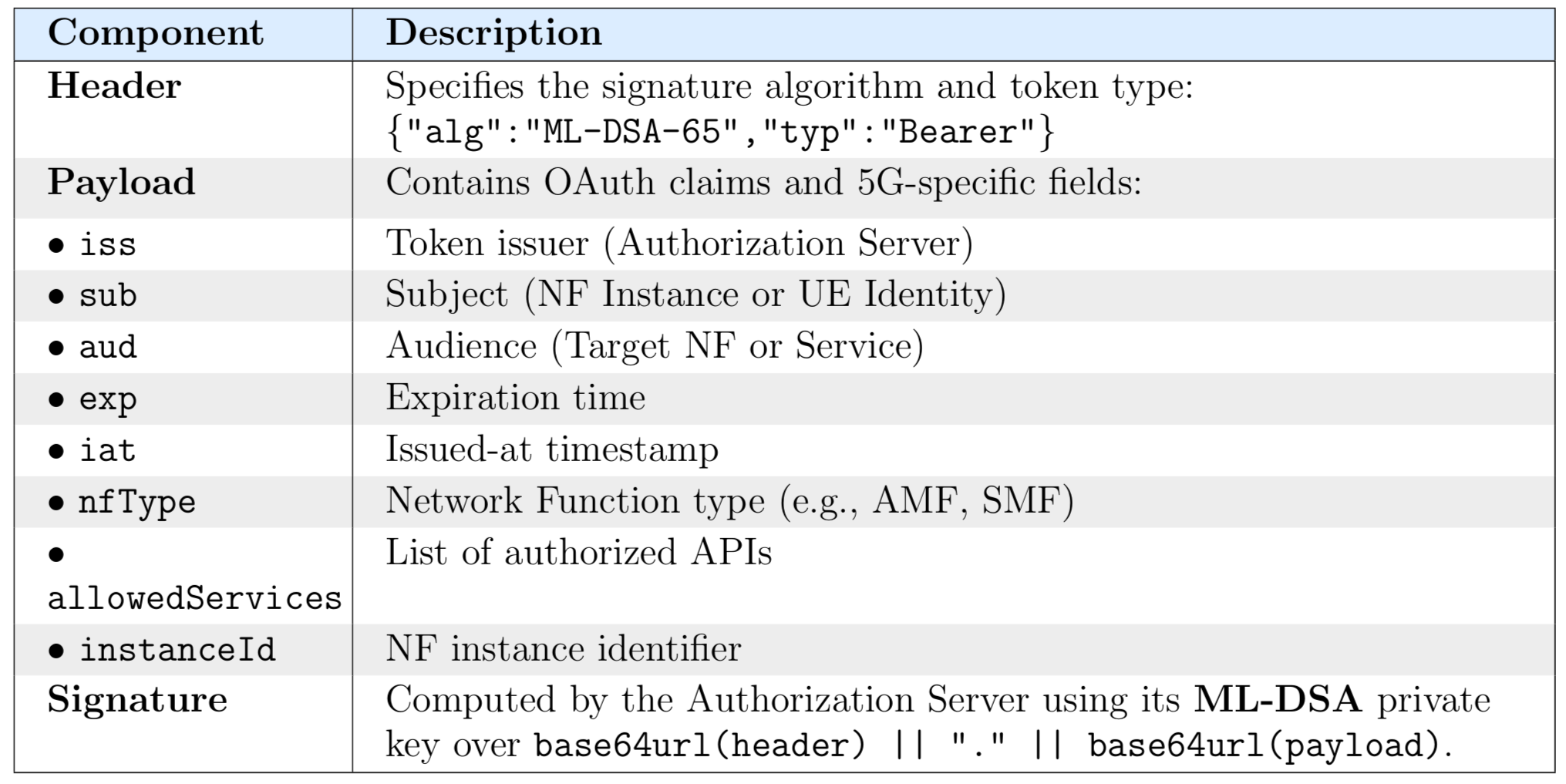}
    \caption{Post-Quantum JWT (PQ-JWT) Token Structure using ML-DSA}
    \label{tab:pqjwt-structure}
\end{figure*}

The structure of the PQ-JWT token is depicted in the table \ref{tab:pqjwt-structure}.




\subsection{PQ-SUCI: Post-Quantum SUCI}

The Subscription Concealed Identifier (SUCI) mechanism in 5G safeguards the permanent subscriber identifier (SUPI) during its transmission between the User Equipment (UE) and the Home Network (HN). To ensure resistance against quantum-capable adversaries, we extend this mechanism to Post-Quantum SUCI (PQ-SUCI). The PQ-SUCI design utilizes ML-KEM-512/768 for forming the initial 32 bytes shared secret, which is then utilized in AES-256 for symmetric encryption, while HMAC-SHA256 provides message integrity checks for the entire protocol flow.

\begin{figure*}
    \centering
    \includegraphics[width=\linewidth]{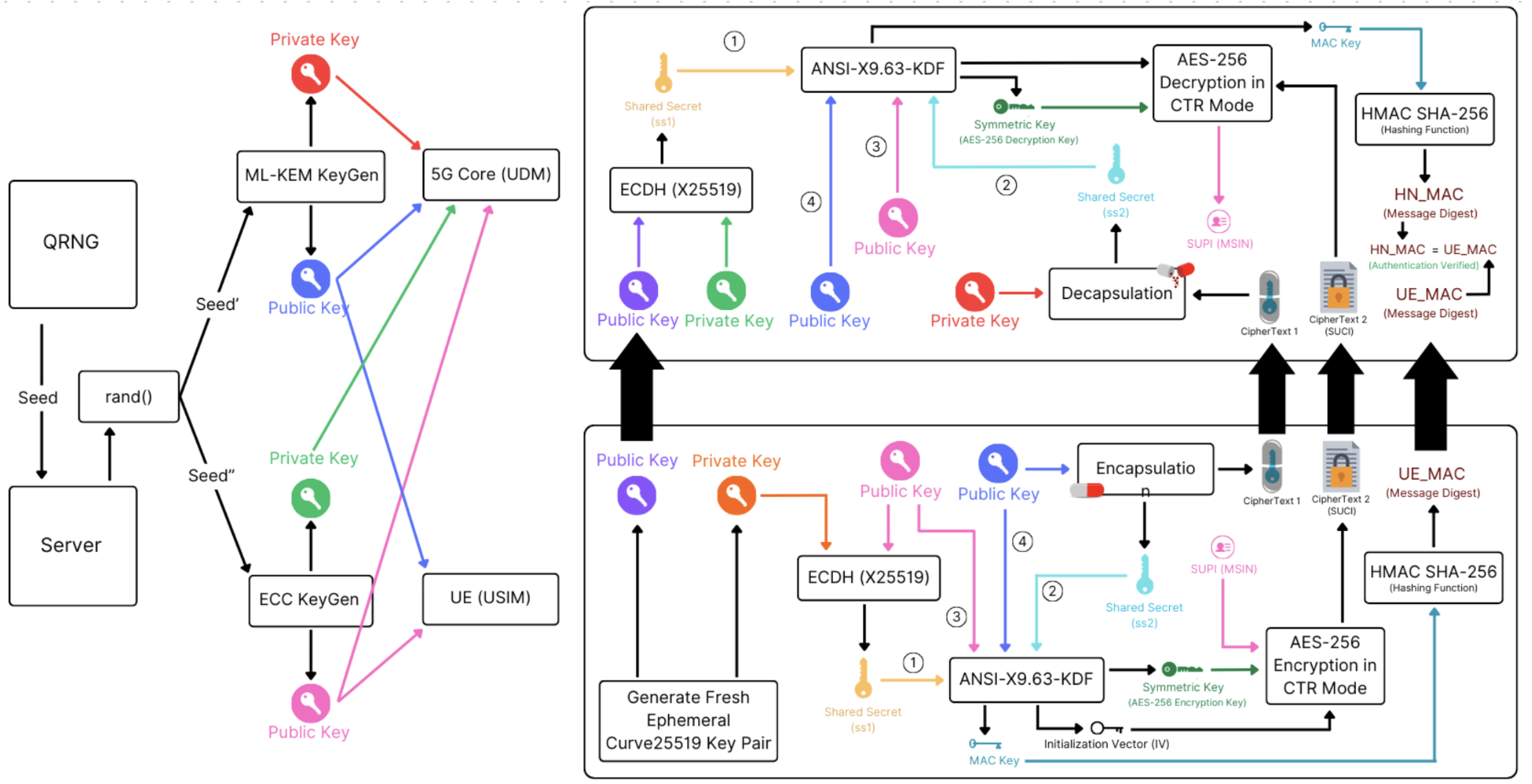}
    \caption{\textbf{Hybrid PQ-SUCI-AKA in 5G}\\The diagram above depicts a hybrid post-quantum protection mechanism for UE subscriber identifiers (SUPI), which incorporates ML-KEM-768 in the ECIES-based key exchange. It utilizes AES-256 in CTR mode with unique nonces to provide symmetric encryption of the SUPI, while HMAC-SHA256 ensures message integrity and authentication. With the desired protocol flow, both the User Equipment (UE) and the Home Network (UDM) will reach identical symmetric and MAC keys via ANSI X9.63-KDF, thus enabling quantum-resilient confidentiality and integrity within 5G authentication procedures.}
    \label{fig:pq-suci}
\end{figure*}

\textbf{Key Generation:} Using the entropy of QRNG, Home Network(HN) generates an ML-KEM key pair. The resulting public key (\textit{hn\_pub\_key}) is then distributed to the UE through the USIM profile, while the private key (\textit{hn\_pvt\_key}) is securely stored in the HN’s Unified Data Management (UDM) component.

\textbf{SUCI Generation:} The UE encapsulates the HN’s public key using ML-KEM to establish a shared secret. The derived secret is expanded via the ANSI X9.63 Key Derivation Function (KDF) to generate symmetric encryption and MAC keys. The SUPI is encrypted using AES-256 in CTR mode, and an HMAC-SHA256 digest (\textit{UE\_MAC}) is computed over the ciphertext to ensure integrity and authenticity. The resulting encrypted identifier and MAC constitute the PQ-SUCI.

\textbf{SUCI Decryption and Verification:} Upon receipt of the PQ-SUCI, the HN decapsulates the ML-KEM ciphertext using its private key to recover the shared secret. The same KDF process regenerates the symmetric and MAC keys. The HN decrypts the SUPI using AES-256 in CTR mode and recomputes the HMAC-SHA256 digest (\textit{HN\_MAC}). Authentication is confirmed if \textit{HN\_MAC = UE\_MAC}, verifying that the SUCI has not been tampered with and originates from a legitimate UE.

\textbf{Security Properties:} The proposed PQ-SUCI mechanism safeguards the confidentiality of the subscriber’s SUPI against both classical and quantum threats. ML-KEM enables PQ secure key encapsulation with Perfect Forward Secrecy (PFS), while AES-256 and HMAC-SHA256 together provide strong symmetric encryption and message integrity. As detailed in the Indian Patent \textit{IN561622: Securing SUCI with Forward Secrecy using Hybrid PQC}~\cite{patent561622}, the hybrid framework combines PQ and classical cryptographic primitives to support crypto-agility and ensure a smooth transition during the PQC migration phase. In doing so, PQ-SUCI mitigates risks such as identity exposure, replay, and impersonation attacks, while remaining fully aligned with 3GPP SUCI specifications and ensuring long-term quantum resilience.

\subsection{Contributions to National Standards and Open-Source Communities}

The Telecommunication Engineering Centre (TEC) Technical Report on Quantum Secure 5G and Beyond 5G Core ~\cite{tec_report}—authored by members of this research team , presents one of the first comprehensive national analyses of the risks of the quantum era facing 5G security architecture. Vulnerabilities across Service-Based Interfaces (SBI), N2, and N3 connections are examined in this report and recommend the adoption of post-quantum cryptography (PQC) mechanisms such as ML-KEM, ML-DSA, SLH-DSA, PQ-TLS/DTLS, and PQ-IPsec to strengthen these interfaces. It further proposes a hybrid migration strategy that combines classical and PQ algorithms, allowing telecom operators to transition securely without operational disruption.

Building on the TEC framework that we co-authored, this paper extends that foundational work into an end-to-end, deployable reference architecture for quantum-resilient 5G Core networks available at {\textbf{\url{https://github.com/coranlabs/QORE}}}. 
The proposed QORE framework translates the national recommendations into a technical implementation blueprint, validated through participation and contributions to open-source 5G projects and the Linux Foundation ~\cite{5g_sbp}. These open-source integrations demonstrate that PQ-TLS, PQ-IPsec, and PQ-OAuth can run effectively in live, virtualized network setups. They also act as reference points that other researchers and developers can build on when exploring post-quantum security in telecom systems.

Overall, this work extends our group’s ongoing involvement in the TEC’s national standardization initiatives and open-source projects focused on post-quantum network security. It aims to close the gap between policy, standards, and real-world deployment.

\subsection{Post-Quantum Security Infrastructure and Technology Stack}

\textbf{Post-Quantum PKI:} We implement a complete post-quantum Public Key Infrastructure supporting ML-DSA certificate issuance, distribution, renewal, and revocation. The root CA uses ML-DSA-87 for maximum security, while intermediate CAs may use ML-DSA-65 for efficiency. Certificate chains maintain quantum resistance throughout the trust hierarchy.

\begin{figure}
    \centering
    \includegraphics[width=\linewidth]{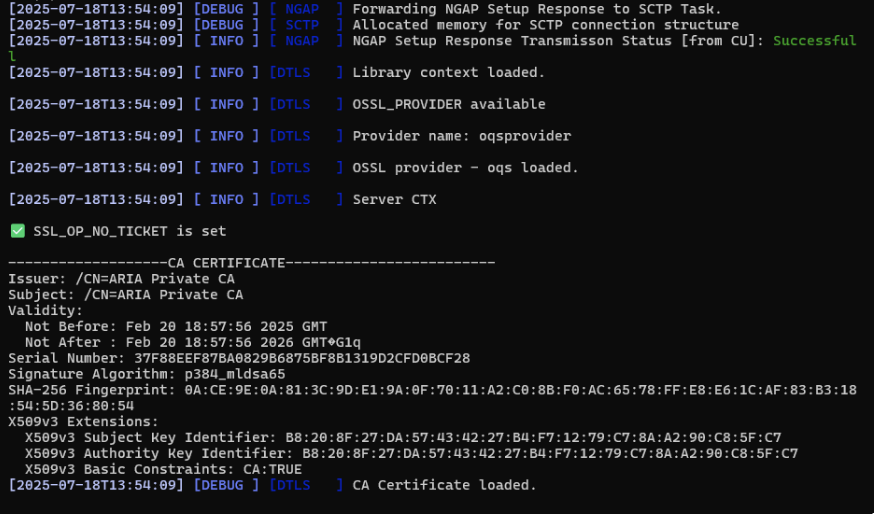}
    \caption{PQ-DTLS Certificate Authority Structure. This diagram illustrates the CA hierarchy for issuing PQ-DTLS certificates to RAN components (gNBs) and core network functions interfacing with RAN. The root CA uses ML-DSA-87 for signing intermediate CA certificates, which then issue end-entity certificates to gNBs using ML-DSA-65. Certificate chains are validated during PQ-DTLS handshake, with explicit handling of fragmentation for large ML-DSA certificates over SCTP transport.}
    \label{fig:pq-dtls}
\end{figure}

\textbf{DTLS Migration:} Figure~\ref{fig:pq-dtls} shows the PQ-DTLS 1.3 implementation for N2 control plane message security. The implementation handles SCTP-specific challenges including larger certificate messages (requiring fragmentation) and retransmission logic for reliability.

\begin{figure*}
    \centering
    \includegraphics[width=\textwidth]{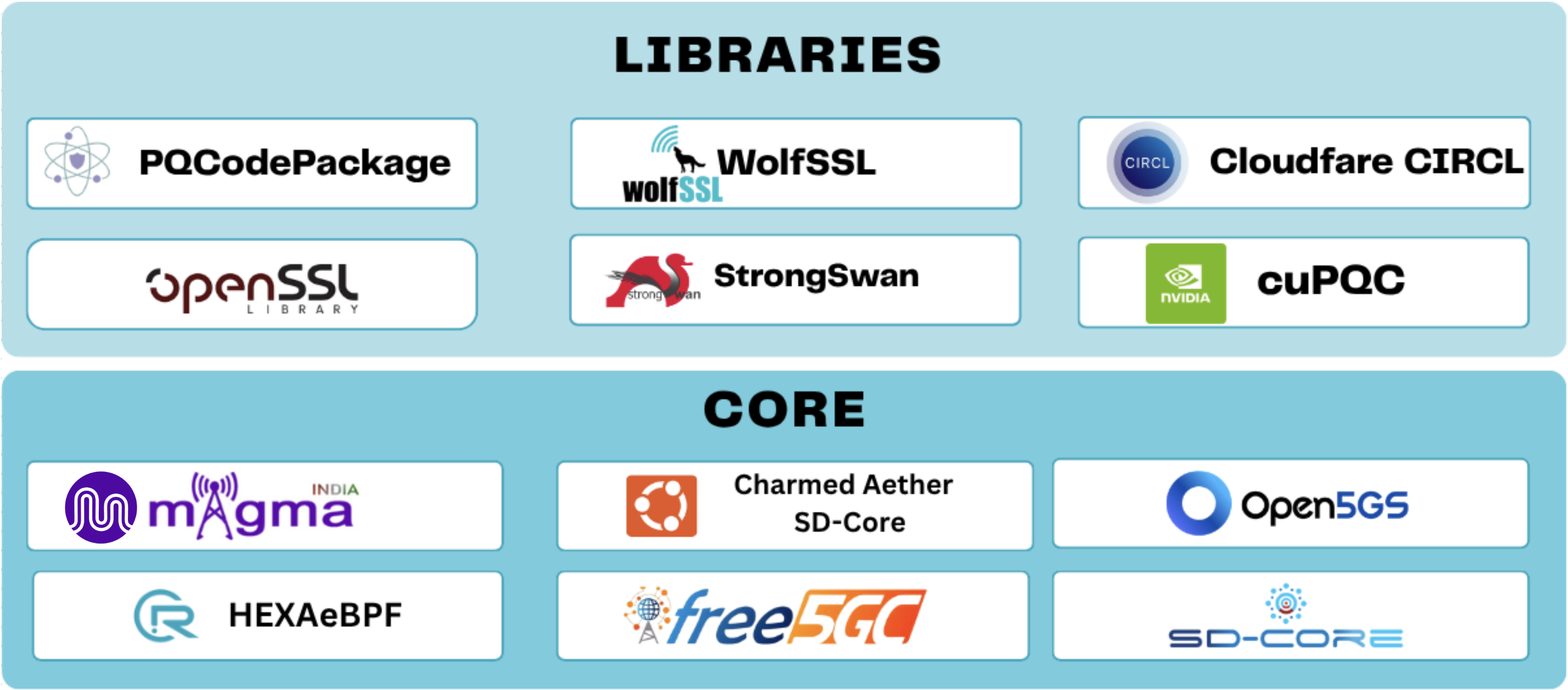}
    \caption{Libraries and Technology Stack for PQ-5GC Implementation. The ecosystem is organized into three layers: Libraries (cryptographic implementations and protocol stacks), Core (5GC network function implementations and testing frameworks), and supporting components (hardware acceleration and development tools). Integration points between layers enable modular development and deployment of post-quantum capabilities across the 5G infrastructure.}
    \label{fig:placeholder}
\end{figure*}

\textbf{Technology Stack:} Figure~\ref{fig:placeholder} (Libraries and Tech Stack) presents the comprehensive software ecosystem enabling our PQ-5GC implementation. The stack includes:
\begin{itemize}
\item \textbf{Cryptographic Libraries:} Cloudflare's CIRCL \cite{circl} was used for integrating  PQ cryptographic primitives into TLS 1.3 and PQ-SUCI. Additionally, OpenSSL \cite{openssl} and wolfSSL \cite{wolfssl} were utilized for adding PQ DTLS 1.3 at the AMF's N2 interface. Strongswan (6.0.0) ~\cite{strongswan} for deploying PQ-IPsec/IKEv2 at N2 and N3 interfaces in conjunction with RAN. 

\item \textbf{Core Network:} Aether SD-Core \cite{sdcore}, free5GC \cite{free5gc} for 5GC implementation and integration testing. 

\item \textbf{Development Tools:} PQCP \cite{PQCA2025mlkemNative}, PQ-Clean \cite{pqclean} for algorithm implementations; Cloudflare CIRCL ~\cite{circl} and BoringCrypto \cite{boringcrypto} for optimized libraries.

\item \textbf{Hardware Acceleration:} Cryptographic operations can be accelerated on GPUs using frameworks like cuPQC  , while CPUs benefit from SIMD-based intrinsics (e.g., AVX2/AVX512)\cite{PQCA2025mlkemNative} to speed up Number Theoretic Transform (NTT) and polynomial operations which are frequently used in lattice-based schemes. Additional optimizations include architecture-specific bitwise instructions (e.g., PCMULQDQ) for Hamming weight computations, coefficient sampling, and signature norm checks. High-throughput implementations can also leverage FPGA or ASIC accelerators for core lattice operations, including NTT \cite{Kundi2024HighPerformanceNTT}, sampling, hashing and matrix-vector multiplications.
\end{itemize}

\textbf{Complete Architecture:} Figure~\ref{fig:pq-5gc} presents the holistic quantum-secure 5G Core architecture, integrating all PQ protocols:

\begin{figure*}
    \centering
    \includegraphics[width=\textwidth]{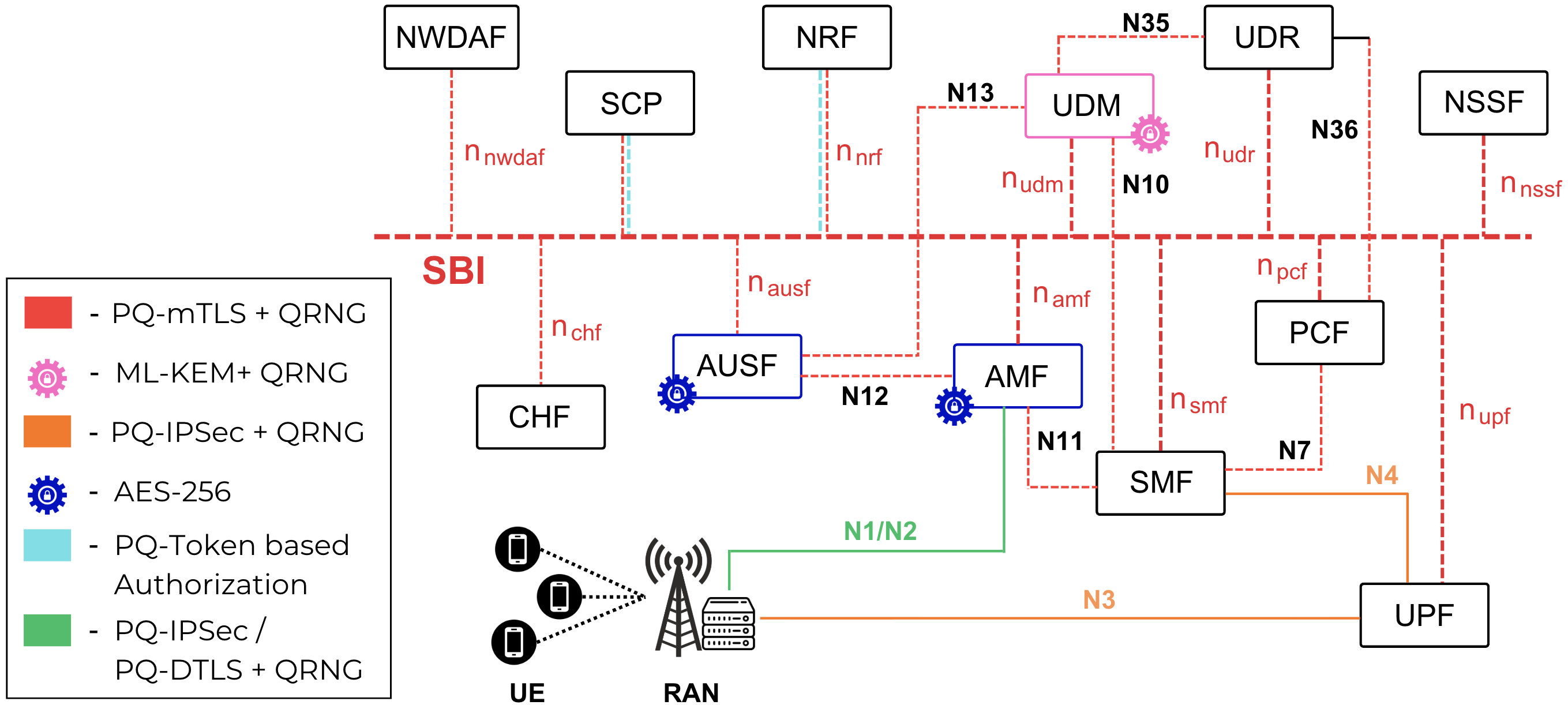}
    \caption{Complete Post-Quantum Secure 5G Core Architecture. This comprehensive diagram shows the integration of all quantum-resistant protocols across the 5G infrastructure: (1) PQ-mTLS securing SBI communications between NFs (AMF, SMF, UPF, AUSF, UDM, NRF), (2) PQ-OAuth for token-based authorization managed by NRF, (3) PQ-IPsec/PQ-DTLS protecting N2 and N3 interfaces between gNB and core, (4) ML-KEM based SUCI encryption for subscriber identity protection, (5) AES-256 symmetric encryption for user data, and (6) integration with QRNG for quantum-safe random number generation. Color coding indicates different security mechanisms: red for mTLS/QRNG, blue for ML-KEM/QRNG, green for IPsec/DTLS, and yellow for OAuth/token authentication.}
    \label{fig:pq-5gc}
\end{figure*}

\section{Performance Evaluation}

The practical deployment of post-quantum cryptography in 5G Core networks requires rigorous performance validation to ensure the algorithms meet the stringent latency, throughput, and resource requirements of telecommunications infrastructure.

\begin{table*}[h!]
\centering
\caption{Post-Quantum Cryptography (PQC) and Classical Algorithm Performance on Intel i9 12th Gen (24 Cores) \& NVIDIA A4000x}
\label{tab:pqc_performance}
\begin{tabular}{lccccc}
\hline
\textbf{Algorithm Name} & \textbf{Keygens/s}& \textbf{Encaps/s} & \textbf{Decaps/s} \\
\hline\\
MLKEM\_512\_cuPQC & 3.16M & 3.22M & 3.10M  \\
mlkem512 & 377,853.3 & 401,310 & 417,237.7  \\
MLKEM\_768\_cuPQC & 2.42M & 2.475M & 1.6M  \\
mlkem768 & 236,076.3 & 259,276 & 261,701.5  \\
mlkem1024 & 173,090.3 & 179,965.2 & 181,064.9  \\
X25519 & 96,750.6 & 47,330.3 & 95,542.3  \\
X448 & 17,081.7 & 8,592 & 17,967.3 &  \\
frodo976aes & 5,348.3 & 3,990 & 4,198.6  \\
bike\_l5 & 2,292.2 & 16,137.3 & 542.8  \\
frodo976shake & 1,522 & 1,380 & 1,416.3  \\
hqc\_192 & 644.8 & 310.1 & 212.6  \\
RSA\_3072 & 18.7 & 78,978 & 1,570.1   \\
\hline
\end{tabular}
\end{table*}

\subsection{Computational Performance}

Table~\ref{tab:pqc_performance} presents benchmark results for key encapsulation mechanisms, comparing ML-KEM variants against classical alternatives and other PQC candidates. The evaluation platform consists of Intel i9 12th Gen processor (24 cores) with NVIDIA A4000 GPU acceleration.

\textbf{Key Findings:}
\begin{itemize}
\item ML-KEM-512 with GPU acceleration (cuPQC) achieves 3.16M key generations per second, demonstrating exceptional performance suitable for high-scale deployments
\item ML-KEM-768, the recommended security level for 5G, maintains high throughput (236K keygens/s, 259K encaps/s) while providing security equivalent to AES-192
\item Classical X25519 performs at 96K keygens/s, demonstrating that ML-KEM-768 provides 2.4× better performance with superior quantum resistance
\item Code-based (HQC) and isogeny-based (BIKE) alternatives show significantly lower performance, validating our choice of lattice-based ML-KEM
\item GPU acceleration provides up to 10× performance improvement for ML-KEM operations, enabling cost-effective scaling for high-throughput scenarios
\end{itemize}

Table~\ref{tab:mldsa_performance} evaluates ML-DSA signature performance, critical for certificate operations and OAuth token signing:

\begin{table*}[h!]
\centering
\caption{Performance of MLDSA Post-Quantum Signature Schemes on Intel i9 12th Gen (24 Cores) \& NVIDIA A4000x}
\label{tab:mldsa_performance}
\begin{tabular}{lcccc}
\hline
\textbf{Algorithm Name} & \textbf{Keygens/s} & \textbf{Signs/s} & \textbf{Verifies/s}  \\
\hline
MLDSA\_44 & 2.14M & 0.25M & 1.15M \\
MLDSA\_65 & 1.32M & 0.20M & 0.82M \\
\hline
\end{tabular}
\end{table*}

\textbf{Signature Performance:}
\begin{itemize}
\item ML-DSA-44 achieves 250K signatures per second, adequate for typical SBI communication rates
\item Verification operations (1.15M/s) significantly outperform signing, ensuring minimal overhead for token validation
\item GPU acceleration enables 2.14M key generations per second, supporting rapid certificate provisioning during network expansion or recovery scenarios
\end{itemize}

\subsection{Operational Impact}

\textbf{Certificate Size Impact:} ML-DSA signatures increase certificate sizes from roughly 1~KB (ECDSA) to about 4--5~KB (ML-DSA-65). Although this results in a higher handshake overhead, the effect is largely mitigated through TLS~1.3 certificate compression and caching mechanisms. In the context of persistent 5GC SBI sessions, the one-time cost of the handshake is amortized across thousands of subsequent service requests.

\textbf{Handshake Latency:} Experimental measurements show that PQ-TLS handshakes introduce an additional latency of approximately 8--12~ms compared to classical TLS, primarily due to the larger certificate payloads. For 5GC control-plane transactions, where the typical RTT ranges from 20--50~ms, this corresponds to a modest 20--30\% increase—an acceptable trade-off considering the substantial security gains. User-plane traffic remains unaffected, as symmetric encryption (AES-256-GCM) is unchanged.

\textbf{Memory Footprint:} Implementations of ML-KEM and ML-DSA add about 30--50~KB of extra memory per connection to maintain cryptographic state, which is negligible in modern virtualized NF environments where multi-gigabyte memory allocations are standard.

\subsection{Scalability Analysis}

The performance evaluation confirms that PQC algorithms, particularly ML-KEM and ML-DSA, satisfy 5G Core performance benchmarks:

\begin{itemize}
\item \textbf{Session Establishment Rate:} A single core can initiate more than 50{,}000 PQ-TLS sessions per second (limited mainly by signing operations), sufficient to accommodate connection surges during network events.
\item \textbf{Authorization Throughput:} OAuth token verification achieves 1.15~million operations per second, supporting millions of concurrent service requests and surpassing typical NRF authorization demands.

\item \textbf{IPsec Throughput:} AES-256-GCM encryption—identical to classical implementations—sustains throughput between 10 and 40~Gbps per core, depending on packet size, thereby meeting N3 user-plane bandwidth requirements.
\end{itemize}

The complete implementation is publicly available for reproducibility and further experimentation at:  
\textbf{\url{https://github.com/coRANlabs/QORE/}}

\section{Migration Strategy and Deployment Considerations}

\subsection{Phased Migration Approach}

\textbf{Phase 1 — Core NF SBI:} Implement PQ-mTLS between key NFs (AMF, SMF, AUSF, UDM, NRF) using hybrid ML-KEM+X25519 key exchange and ML-DSA-based certificates. Retain classical TLS as a fallback option to ensure interoperability during the transition phase.

\textbf{Phase 2 — RAN Interfaces:} Upgrade N2/N3 IPsec tunnels to PQ-IPsec with Pre-Shared Public Key (PPK) authentication, and enable PQ-DTLS for securing N2 control-plane traffic. This phase requires coordination with RAN vendors for gNB software updates to integrate PQ-enabled cryptographic modules.

\textbf{Phase 3 — OAuth and UE Security:} Introduce PQ-OAuth using ML-DSA-signed tokens to strengthen authentication. Apply ML-KEM-based SUCI encryption to protect subscriber identities, with corresponding UE and USIM firmware updates carried out in collaboration with device manufacturers.

\textbf{Phase 4 — Optimization and Homogeneous PQ:} Decommission classical fallback mechanisms once PQC solutions reach operational maturity. Refine PQC rollouts with insights gathered from field performances and integrate PQ-relevant hardware accelerators for speedup wherever advantageous.

\subsection{Interoperability and Standards}

The implementation we presented aligns closely with the ongoing global efforts to standardize post-quantum cryptography across networks and within 5G infrastructure. Collectively, these initiatives have a common goal, i.e., to ensure safer, future-proof, operational, and performant networks, thereby providing ease of operations both to the user and the operators. The hybrid approach enables a seamless transition to quantum safety across diverse telecom networks.

\begin{itemize}

\item \textbf{3GPP SA3 PQC Study Items:}
The 3GPP SA3 working group has undertaken detailed evaluations to assess the feasibility of integrating PQC mechanisms into the 5G security framework. These studies focus on defining interfaces between Access and Core Networks using PQC primitives to ensure they can be incorporated without disrupting the existing security mechanisms ~\cite{3GPP_SA3_PQC_5G}. Furthermore, they focus on key management, authentication, and signaling protection using the standard post-quantum primitives.

\item \textbf{IETF RFC 9370 and 8784:}
RFC~9370 ~\cite{rfc9370} and 8784 ~\cite{rfc8784} are significant steps towards achieving quantum safety at the Internet Protocol Layer, making it highly useful for IP-dependent networks like the 5G Core. The integration of 256-bit pre-shared keys and the addition of new messages for carrying out post-quantum key exchange allow for a gradual transition, which is necessary for maintaining interoperability, backward compatibility with legacy peers, and trust.

\item \textbf{NIST PQC Standardization:}
NIST’s PQC standardization program (FIPS~203/204/205) defines new quantum-safe cryptographic primitives, namely---ML-KEM, ML-DSA, and SLH-DSA. These algorithms serve as the foundation for the proposed post-quantum upgrades of classical protocols, providing a balance between efficiency and security. Additional recommendations from NIST, such as ~\cite{nist_kem_rec}, provide the techniques and guidance to deploy these algorithms, with the best possible security guarantees without hampering performance. This serves as a reference framework for telecom-grade PQC standardization efforts.

\item \textbf{Ongoing IETF Work on PQ-TLS:}
The IETF and IRTF continue to expand the scope of post-quantum cryptography by extending the classical security protocols (e.g., TLS, IPsec, MLS, X.509) with PQC primitives with multiple RFCs and draft specifications in progress. Their hybrid approach to PQC serves as a reference for telecom standard bodies to carry out PQC migration for 5G Infrastructure~\cite{draft-ietf-tls-ecdhe-mlkem,draft-ietf-tls-mldsa,connolly-cfrg-xwing-kem,draft-ietf-lamps-dilithium-certificates}.


\item \textbf{GSMA PQ.05 -- PQ.03 Post Quantum Cryptography Guidelines for Telecom Use Cases v2.0 (2025):}
The GSMA PQ.05 whitepaper ~\cite{gsma_pq_guidelines} serves as a complete guide detailing the Post-Quantum migration for Telecom operators, starting from the planning phase, implications, summaries, use-cases, and migration scenarios. It frequently refers to the PQ migration recommendations issued by the respective institutes of countries, e.g., the UK, Australia, the U.S., etc, to offer verified directions. The whitepaper frequently highlights the mentioned points:
    \begin{itemize}
        \item Lists the migration options, the cryptographic primitives (asymmetric, symmetric) affected by the emergence of quantum computers. 
        \item Impacts on classical protocols (TLS, IPsec).
        \item Zero-Trust Architecture using PQC primitives
        \item Crypto-agility, migration strategies, roadmaps for Mobile Network Operators (MNOs). Additionally, it highlights the implications for PKI.
        \item Scope of PQC in Telecom, delving into topics like Base Stations, Virtual Network Functions, SIM Provisioning, etc.
    \end{itemize}

\item \textbf{GSMA PQ.05 Post-Quantum Cryptography for 5G Roaming use case (2025):}
GSMA PQ.05 Post-Quantum Cryptography for 5G Roaming use case~\cite{gsma_pq05} explores the addition of PQC in securing 5G/4G's inter-PLMN and roaming interfaces, using Hybrid PQC in protocols like TLS, and IKE. The report is a continuation of GSMA's series of documents on Post-Quantum Telco Network (PQTN). The report lays down the following ideas and points:
\begin{itemize}
    \item An overview of the effects of CRQCs on 3GPP-standardized roaming interfaces in both 4G and 5G. This is the scenario where a UE roams between a Home PLMN (HPLMN) to a visiting PLMN (VPLMN). 
    \item A quantum-safe protection scenario for the N32 (SEPP) inter-operator interface, which is a common attack entry-point on an operator's network.
    \item Upgradation for the Protocol for N32 Interconnect Security (PRINS) to quantum-safety using Hybrid TLS and PQ X.509 mechanisms for N32-c/N32-f. It also prescribes the use of JSON Web Encryption, with an AEAD extension, and JSON Web Signatures (JWS) upgraded using PQ Signature schemes.
    \item The report also focuses on 4G roaming security, where Diameter messages are encrypted using PQ-IPsec/PQ-DTLS utilizing ML-KEM as the key exchange algorithm. Additionally, quantum-safe encryption for VoLTE and VoNR has been mentioned, focusing on IPsec and IMS AKA.
    \item It concludes by describing scope, cryptographic inventories, etc.
    \end{itemize}
\end{itemize}

Active participation in cross-industry groups such as 5G-ACIA and GSMA is essential for collective visions, multi-vendor interoperability, and for defining an ecosystem that is fully prepared to deal with the quantum era. The standards harmonization rendered by the joint collaboration of these groups drives 5G networks forward.

\subsection{Operational Challenges}

\textbf{Key Management Complexity:} 
The integration of PQC adds different key types with different workflows. The PKI infrastructure needs to be configured to support the full lifecycle of ML-DSA Certificates while managing shared public key(PPK) distribution for IPsec and ensuring that the storage and rotation of all associated key takes place securely and efficiently.

\textbf{Certificate Distribution:} 
With robust security comes an overhead of large sizes of certificates which pose challenges for bandwidth-constrained or environments with limited resources. Implementation of certificate compression, caching and efficient distribution mechanisms will help minimize the impact of large sized certificates on management and control networks.

\textbf{Algorithm Agility:} 
A secure transition is made possible by the hybrid deployment model, but operational complexity is also increased. In order to eventually simplify configurations and preserve long-term cryptographic simplicity, it is imperative to establish explicit deprecation timelines for classical algorithms.

\textbf{Training and Expertise:} 
Network operators and security teams must increase their capacity in order for PQC to be adopted successfully. PQC basics, configuration and troubleshooting procedures, and the unique operational considerations related to quantum-resistant cryptography should be the main topics of training.

\section{Future Scope}

Future research and development efforts will focus on operationalizing the quantum-resilience framework across the entire end-to-end 5G system, addressing logistical, cryptographic, and performance challenges beyond the Core.

\subsection{Quantum Key Distribution (QKD) Integration}

While this paper focused on PQC, future work should explore the seamless integration of Quantum Key Distribution (QKD) as a complementary layer to PQC. QKD establishes theoretically unbreakable symmetric keys based on physics, offering a compelling long-term solution for point-to-point secure links, such as high-capacity backbone interfaces. It could help in establishing forward secrecy in long-lived connections without the need to carry out frequent asymmetric rekeying. Research is required to integrate QKD management systems with the existing 5GC key management infrastructure to ensure centralized, crypto-agile operation.

\subsection{Quantum-Resilient RAN (QRAN) Architecture}

The architectural extensions detailed herein for the 5GC must be matched by a corresponding effort to develop a Quantum-Resilient Radio Access Network (QRAN) architecture. This involves extending PQC protection, specifically the Hybrid IKEv2 Key Exchange, to the N2/N3 interfaces between the gNBs and the Core and also between the components of the RAN infrastructure.

\subsection{PQ-SUPI Testing on Real User Equipment (UE)}

Following the architectural definition of the quantum-safe SUPI-to-SUCI conversion, empirical testing on actual User Equipment (UE) environments is necessary. This involves implementing the Hybrid and Homogeneous ML-KEM solutions for subscriber identity concealment and measuring performance metrics such as latency and power consumption on real UE platforms and their USIMs, validating the efficiency and security of the proposed PQC approach in an end-to-end context.

\subsection{AES Standard Upgradation for Ciphering and Integrity Protection}

Alongside PQC adoption and in accordance with 3GPP~TS~35.245~\cite{3gpp_35245}, a key enhancement involves upgrading the Advanced Encryption Standard (AES) from 128-bit to 256-bit key lengths across both Core and User Equipment (UE) domains. Transitioning to AES-256 strengthens symmetric encryption against brute-force and quantum-assisted attacks. Future 5G and 6G security profiles are expected to adopt AES-256 for both ciphering and integrity protection, replacing the existing NIA2 and NEA2 algorithms with their AES-256–based counterparts (NIA5/NEA5)~\cite{3gpp_35245}. This upgrade ensures end-to-end quantum resilience, preserving confidentiality and integrity even in the presence of advanced quantum-capable adversaries.

\subsection{Core Network Implementations}

As illustrated in Figure~\ref{fig:placeholder}, several open-source 5G Core implementations are currently available, including \textbf{Magma}~\cite{magma}, \textbf{Aether}~\cite{aether}, \textbf{HEXA eBPF} ~\cite{hexa_ebpf}, \textbf{SD-Core}~\cite{sdcore}, \textbf{free5GC}~\cite{free5gc}, \textbf{Charmed Aether SD-Core}, and \textbf{Open5GS}~\cite{open5gs}.

In this work, we deployed and validated the proposed QORE framework on open-source 5G Core platforms such as free5GC and Aether, demonstrating its adaptability and confirming that QORE can function as a reference architecture deployable across both open-source and production-grade 5G Core systems. A demonstration of the PQC-enhanced Charmed Aether SD-Core implementation is available at \href{https://youtu.be/_yWoZEG-8Go?si=In_QxvK6K5Q_DjNR}{YouTube}. To perform comparative performance studies and verify interoperability across various 5GC environments, we plan to expand our evaluations to other platforms in the future, including \textbf{Magma}, \textbf{OAI}, and \textbf{Open5GS}.

\section{Conclusion}

The work introduced \textbf{QORE}, a practical framework demonstrating the integration of post-quantum cryptography into open-source 5G Cores, future-proofing them against the emerging quantum threats to classical cryptographic security protocols. \textbf{QORE} provides a consistent and seamless upgrade from classical security protocols to quantum-resistant protocols, simultaneously sustaining desired network architecture and performance.

The contributions of this research play an important role in the field of security in 5G systems. QORE establishes itself as a reference point from which other network operators or researchers can use a blueprint when planning their own quantum-safe migrations and upgrades. Prior to production deployments, operators or researchers can test strategies, measure performance impact, and validate security assumptions using QORE, which is more than just a quantum secure template.

In this  research  we explain in-depth how critical security  protocols   such as TLS, IPsec, mTLS and OAuth2. 0 can be redesigned by replacing quantum vulnerable primitives such as RSA and ECC with NIST standardized ML-KEM ( for key encapsulation ) and ML-DSA ( for digital signatures ) algorithms. We have carried out a hybrid PQC model in  QORE  which not only provides a smooth transition but also maintains backwards compatibility to the existing 5G  systems  ensuring interoperability,  trust  and testing of performance. 

Lattice-based PQC algorithms can meet or occasionally even surpass the uncompromising performance requirements of carrier grade infrastructure, according to our extensive evaluations. While ML-DSA uses the NVIDIA cuPQC to verify over 1,000,000 signatures per second on GPU platforms, ML-KEM attains key exchange rates of over 200,000 operations per second. It was observed that the CPU implementations of these algorithms invariably match or at times even surpass the performance of classical  algorithms  namely X25519. Hardware acceleration on the N2 interface leverages NVIDIA Data Processing Units  (DPUs) ,  specifically  \textit{\textbf{Bluefield-3}}  to offload the operations in the IPsec and kTLS protocols with the help of DOCA  libraries ,  which reduce CPU load and improve the overall performance and throughput. These measurements also serve as an important piece of evidence that deployment of PQC within real 5G environments is not only technically feasible but introduces only a marginal latency overhead relative to classical cryptography.

This quantum-proof prototype directly addresses the growing risk of  Harvest  Now, Decrypt  Later,  which protects long-lived sensitive data. These include subscriber identifiers, authentication materials, and signaling  data  such as subscriber identifiers (SUPI, SUCI, GUTI, etc.), authentication materials (long-term keys, AVs), and signaling data that adversaries might collect and store now for future decryption. With the use of quantum-resistant cryptographic techniques across vulnerable network interfaces like N2 and N3, SBI communications, and internal mechanisms for authentication, we provide a reconcilable framework across the entire 5GC architecture, hence providing a robust and long-lasting system.

As we move further, we recognize that sustained collaboration between  operators ,  equipment providing vendors, standards  bodies  and the broader research community will have to come together for the large-scale adoption of QORE, representing a crucial step toward a quantum-resilient telecommunications ecosystem. The phased migration plan we described in this work offers a realistic route toward quantum-resilient 5G operations, one that aligns with ongoing  3GPP~SA3  initiatives and NIST PQC recommendations.

\section*{Acknowledgments}

The authors of the paper thank \textbf{coRAN Labs}, \textbf{Netweb Technologies}, and \textbf{Synergy Quantum} for their ongoing help and support throughout the whole project. We also want to thank the 5G research community that works on open-source projects. Lastly, we want to thank the NIST PQC standardization program for its work, which has been very helpful in moving forward research and use of post-quantum cryptography.

\bibliographystyle{IEEEtran}
\bibliography{reference}

\end{document}